\documentclass{eptcs}

\usepackage[ddmmyy,hhmmss]{datetime}
\usepackage{amsthm}
\usepackage{amssymb}
\usepackage{graphics,latexsym,amsfonts}
\usepackage{mathtools}
\usepackage{epigraph}
\usepackage{picture}
\usepackage{enumitem}
\setlist[enumerate]{leftmargin=*}
\usepackage{color}
\usepackage{xspace}
\usepackage{setspace} 

\newtheorem{theorem}{Theorem} 
\newtheorem{definition}{Definition}
\newtheorem{lemma}{Lemma}

\newtheorem{example}{Example}
\newtheorem{claim}{Claim}
\newtheorem{remark}{Remark}

\newcommand{\NP}{NP}
\newcommand{\NEXP}{NEXPTIME}
\newcommand{\terms}{\mathcal{T}_{\varphi}}

\newcommand{\Mi}{}

\newcommand{\medcup}{\textstyle\bigcup\displaystyle}\newcommand{\medcap}{\textstyle\bigcap\displaystyle}

\def\eod {{\unskip\nobreak\hfil\penalty50
\hskip2em\hbox{}\nobreak\hfil \rule{2.57mm}{2,57mm}\hspace{.71pt}
\parfillskip=0pt \finalhyphendemerits=0 \par \medskip}}

\author{
Domenico Cantone\thanks{Department of Mathematics and Computer Science, University of Catania, Italy. Email: domenico.cantone@unict.it}$\:$,
Alfio Giarlotta\thanks{Department of Economics and Business, University of Catania, Italy. Email: alfio.giarlotta@unict.it}$\:$,
Pietro Maugeri\thanks{Department of Mathematics and Computer Science, University of Catania, Italy. Email: pietro.maugeri@unict.it}$\:$,
Stephen Watson\thanks{Department of Mathematics and Statistics, York University, Toronto, Canada. Email: watson@mathstat.yorku.ca}}

\title{\bf Extending rational choice behavior:\\ The decision problem for {B}oolean set theory with\\ a choice correspondence}

\newcommand{\defAs}{\coloneqq}

\newcommand{\pow}{\mathrm{Pow}}

\newcommand{\BSTC}{\mathsf{BSTC}}

\newcommand{\model}{\ensuremath{\mbox{\boldmath $\mathcal{M}$}}\xspace}
\newcommand{\M}[1]{#1^{\scriptscriptstyle M}}

\newcommand{\Mbeta}[1]{#1^{\scriptscriptstyle M_{\beta}}}
\newcommand{\Malpha}[1]{#1^{\scriptscriptstyle M_{\alpha}}}

\newcommand{\choice}{\mathtt{c}}

\newcommand{\true}{\ensuremath{\mbox{$\mathsf{1}$}}\xspace}
\newcommand{\false}{\ensuremath{\mbox{$\mathsf{0}$}}\xspace}
\newcommand{\powPlus}{\pow^{\hbox{\tiny{+}}}(U)}
\newcommand{\powpowPlus}{\pow^{\hbox{\tiny{+}}}\big(\pow^{\hbox{\tiny{+}}}(U)\big)}

\newcommand{\omQprime}{\overline{\mQ}\raisebox{0.6pt}{$'$}}
\newcommand{\omQ}{\overline{\mQ}}

\def\F{\mathcal{F}}

\begin{document}

\maketitle

\begin{abstract}
\noindent 
Given the family $\powPlus$ of all nonempty subsets of a set $U$ of alternatives, a choice over $U$ is a function $c \colon \Omega \to \powPlus$ such that $\Omega \subseteq \powPlus$ and $c(B) \subseteq B$ for all menus $B \in \Omega$.  A choice is total if $\Omega = \powPlus$, and partial otherwise. 
In economics, an agent is considered rational whenever her choice behavior satisfies suitable axioms of consistency, which are properties quantified over menus.    
Here we address the following lifting problem: Given a partial choice satisfying one or more axioms of consistency, is it possible to extend it to a total choice satisfying the same axioms?
We characterize the lifting of some choice properties that are well-known in the economics literature.  
In addition, we study the decidability of the connected satisfiability problem for unquantified formulae of an elementary fragment of set theory, which involves a choice function symbol, the Boolean set operators, the singleton, the equality and inclusion predicates, and the propositional connectives. 
In two cases we prove that the satisfiability problem is \NP-complete, whereas in the remaining cases we obtain \NP-completeness under the additional assumption that the number of choice terms is constant.
\medskip


\end{abstract}

\section{Introduction} \label{SECT:intro}

In this paper we study some theoretical and computational problems that are relevant to \textit{rational choice theory}, which is a widespread framework used to model economic behavior. 
Formally, a \textit{choice} over a ground set $U$ of alternatives is a correspondence $B \mapsto c(B)$, which associates to each `feasible' menu $B \subseteq U$ a nonempty choice set $c(B) \subseteq B$: here each choice set comprises all elements (called items) of the corresponding menu that are deemed selectable by the economic agent.  
A choice can be either \textit{total} (i.e, defined for all nonempty subsets of $U$) or \textit{partial} (i.e., undefined for some, but not all, subsets of $U$).

According to the \textit{Theory of Revealed Preferences} pioneered by the economist Paul Samuelson~\cite{Sam38}, consumers' preferences can be derived from their purchasing habits: in a nutshell, the agent's choice behavior is observed, and his underlying preference structure is inferred. 
The preference revealed by an observed choice is typically modeled by means of a binary relation on $U$.
Then a choice is said to be \textit{rationalizable} when the agent's observed behavior can be univocally retrieved by maximizing his binary relation of revealed preference.

Since the seminal paper of Samuelson, a lot of attention has been devoted to notions of rationality within the framework of choice theory: see, among the plethora of contributions to the topic, the classical papers~\cite{Arr59,Han68,Hou50,Ric66,Sen71}.\footnote{See also the book~\cite{AleBouMon07} for an analysis of the links among the theories of choice, preference, and utility represnetations, as well as the recent survey~\cite{Gia19} for an account of the last trends in preference modeling and boundedly rational choices.} Classically, the rationality of an observed choice behavior is connected to the satisfaction of suitable \textit{axioms of choice consistency}: these are rules of selections of items within menus, which are codified by sentences of second-order monadic logic, universally quantified over menus and items.
Among the several axioms introduced in the specialized literature, let us recall the following:
\begin{itemize}
	\item \emph{standard contraction consistency} $(\alpha)$, introduced by Chernoff~\cite{Che54};
	\item \textit{standard expansion consistency} $(\gamma)$, and \textit{binary expansion consistency} $(\beta)$, both due to Sen~\cite{Sen71};
	\item \textit{the weak axiom of revealed preference} (\textsf{WARP}), due to Samuelson~\cite{Sam38}.
\end{itemize}
It is well-known that, under (rather weak) assumptions on the domain, a choice is rationalizable if and only if the two consistency axioms $(\alpha)$ and $(\gamma)$ hold. 
In addition, this rationalizing preference satisfies the transitive property if and only if axioms $(\alpha)$ and $(\beta)$ hold if and only if \textsf{WARP} holds: in this case, we speak of a \textit{transitively rationalizable} choice. 

Rationalizability is often considered too strong an assumption to be satisfied by the choice behavior of an economic agent. 
Thus,  some other axioms of choice consistency are considered in the literature. 
In this paper we shall analyze the following:
\begin{itemize}
	\item \emph{path independence} ($\textsf{PI}$), due to Plott~\cite{Plott73};
	\item \emph{standard replacement consistency} $(\rho)$, recently introduced in~\cite{CanGiaGreWat16}.
\end{itemize}
These axioms, which yield weaker forms of choice rationality, are surprisingly related to some combinatorial areas of mathematics, such as \textit{convex geometries}~\cite{EdeJam85}.  
Section~\ref{SECT:preliminaries} shall provide all the necessary background to choice theory.

Although the economics literature in rational choice theory is astonishingly large, we were unable to find contributions that explicitly deal with logical aspects of the associated decision procedures.  
In this paper we start filling this gap.
Specifically, here we shall address the following \emph{lifting problem}: Given a partial choice satisfying some axioms of consistency, can we characterize whether it is extendable to a total choice satisfying the same axioms? 
The lifting problem for various combinations of axioms $(\alpha)$, $(\gamma$), $(\beta)$, and $(\rho)$ will be addressed in depth in Section~\ref{SECT:liftings}.

On a more technical register, here we also study the associated satisfiability problem for unquantified formulae of an elementary fragment of set theory (denoted $\BSTC$, for Boolean Set Theory with a Choice operator) involving a choice function symbol $\choice$, the Boolean set operators $\cup$, $\cap$ and $\setminus$, the singleton $\{\cdot\}$, the predicates equality $=$ and inclusion $\subseteq$, and the propositional connectives $\land$, $\lor$, $\neg$, $\implies$, etc. 
We consider the cases in which the interpretation of $\choice$ is subject to suitable combinations of the axioms of choice consistency mentioned above.  
In two cases we prove that the related satisfiability problem is \NP-complete, whereas in some other cases we obtain \NP-completeness only under the additional assumption that the number of choice terms is constant. 

By depriving the $\BSTC$-language of the choice function symbol $\choice$, we obtain the fragment \textsf{2LSS} (here denoted $\BSTC^{-}$), whose decidability was known since the birth of \textit{Computable Set Theory} in the late 70's.\footnote{For an extensive account of this topic, see the monographs \cite{CanFerOmo89a,CanOmoPol01,SchCanOmo11,CanUrs18}.} 
In Section~\ref{appendixDecProc} we rediscover the latter result as a by-product of the solution to the satisfiability problem of $\BSTC$ under the \textsf{WARP}-semantics, by an analysis based on a novel term-oriented non-clausal approach. 
In the case of finite choice correspondences -- which are the ones most considered in economics for their applicability to concrete scenarios --, our solutions to the lifting problem turn out to be effective and -- {\Mi in all cases but one} -- expressible in the $\BSTC$-language. 
This facilitates the design of effective procedures for the solution of the satisfiability problems of our concern. 
The syntax and semantics of the $\BSTC$-language, as well as the solutions of the satisfiability problem for $\BSTC$-formulae under the various combinations of axioms of choice consistency are presented in Section~\ref{SECT:satProb}. 

Finally, in Section~\ref{SECT:Conclusions} we draw our conclusions and hint at future developments.

\section{Preliminaries on choice theory} \label{SECT:preliminaries}

Hereafter, we fix a nonempty set $U$ (the ``universe").
We denote by $\pow(U)$ the family of all subsets of $U$, and by $\powPlus$ its subfamily $\pow(U) \setminus \{\emptyset\}$.
The next definition collects some basic notions in choice theory.

\newcommand{\cPlus}{c^{\hbox{\tiny{+}}}}
\newcommand{\oc}{\overline{c}}
\newcommand{\ocPlus}{\overline{c}^{\hbox{\tiny{+}}}}
\newcommand{\ocPrime}{\overline{\cPlus}}

\begin{definition} \rm \label{DEF:preliminary deff on choice}
    Let $\Omega \subseteq \powPlus$ be nonempty.
    A map $f \colon \Omega \to \pow(U)$ is \textit{contractive} if $f(B) \subseteq B$ for each $B \in \Omega$. 
    A \textit{choice correspondence} on $U$ is a contractive map that is never empty-valued, i.e.,
    $$
    c \colon \Omega \to \powPlus \quad \hbox{such that} \quad c(B) \subseteq B \quad \hbox{for each} \;\; B \in \Omega\,.
    $$
In this paper, we denote a choice correspondence on $U$ by $c \colon \Omega \rightrightarrows U$, and simply refer to it as a \textit{choice}. 
    The family $\Omega$ is the \textit{domain} of $c$, sets in $\Omega$ are \textit{(feasible) menus}, and elements of a menu are \textit{items}. 
    Further, we say that $c \colon \Omega \rightrightarrows U$ is \textit{total} if $\Omega = \powPlus$, and \textit{partial} otherwise.
    The \textit{rejection map} associated to $c$ is the contractive function $\oc \colon \Omega \to \pow(U)$ defined by $\oc(B) := B \setminus c(B)$ for all $B \in \Omega$. \eod
\end{definition}

Note that the rejection map associated to a choice is not, in general, a choice, because it may be empty-valued for some menus. 
Given a choice $c\colon \Omega \rightrightarrows U$, the \textit{choice set} $c(B)$ of a menu $B$ collects the elements of $B$ that are deemed selectable by the economic agent.
Thus, in case $c(B)$ contains more than one element, the selection of a single element of $B$ is deferred to a later time, usually with a different procedure (according to additional information or ``subjective randomization", e.g., flipping a coin).

The next definition recalls the classical notions of revealed preference and rationalizable choice.

\begin{definition} \rm \label{DEF:rationalizable choice}
    Let $c \colon \Omega \rightrightarrows U$ be a choice.
    The \textit{preference revealed by $c$} is the binary relation $\precsim_c$ on $U$ defined by $a \precsim_c b$ if there is a menu $B \in \Omega$ such that $a,b \in B$ and $b \in c(B)$. 
    Then $c$ is \textit{rationalizable} if the equality $c(B) = \max(B,\precsim_c)$ holds for all menus $B \in \Omega$.\footnote{Here $\max(B, \precsim) := \{a \in B : (\nexists b \in B) (a \prec b)\}$ is the set of maximal items of $B$, where $a \prec b$ means $a \precsim b$ and $\neg(b \precsim a)$.} 
    Equivalently,\footnote{\label{foot:asymmetric}This equivalence readily follows from the fact that only the asymmetric part $\prec_c$ of the relation of revealed preference determines the rationalizability of $c$.} $c$ is rationalizable if  there exists a binary relation $\precsim$ on $U$ such that $c(B) = \max(B,\precsim)$ for all $B \in \Omega$.
    \eod
\end{definition}

The approach based on revealed preference theory postulates that preferences can be derived from choices. 
Then a choice is rationalizable whenever the observed behavior can be explained (i.e., retrieved) by constructing the binary relation of revealed preference, and selecting all and only the maximal elements from each each menu.

The rationalizability of a choice is connected to the satisfaction of suitable \textit{axioms of choice consistency}, which encode rules of coherent economic behavior.
Among the several axioms that are considered in the literature, the following are relevant to our analysis (a universal quantification over all the involved menus is implicit): \textbf{(Alfio: I am not sure that axiom $(\rho)$ can be formulated in this way, because in the paper where the equivalence is stated (and not proved!), namely~\cite{CanGiaWat19} p.\,287 (see also Footnote 36), the choice domain is always $\powPlus$.)}
\smallskip

\noindent 
\begin{tabular}{lll}
  \textbf{axiom $(\alpha)$} [\emph{standard contraction consistency}]: & 
      $A \subseteq B \;\; \Longrightarrow \;\; A \cap c(B) \subseteq c(A)$\\[.1cm]
  \textbf{axiom $(\gamma)$} [\emph{standard expansion consistency}]: &
      $c(A) \cap c(B) \subseteq c(A \cup B)$\\[.1cm]
  \textbf{axiom $(\beta)$} [\emph{symmetric expansion consistency}]: &
      $\big(A \subseteq B \: \wedge \: c(A) \cap c(B) \neq \emptyset \big) \;\; \Longrightarrow \;\; c(A) \subseteq c(B)$\\[.1cm]
  {\Mi \textbf{axiom $(\rho)$} [\emph{standard replacement consistency}]:} &
      $c(A) \setminus c(A \cup B) \neq \emptyset \;\; \Longrightarrow \;\; B \cap c(A \cup B) \neq \emptyset$\\[.1cm]
  \textbf{\textsf{WARP}} [\emph{weak axiom of revealed preference}]: &
      $\big(A \subseteq B \: \wedge \: A \cap c(B) \neq \emptyset \big) \;\; \Longrightarrow \;\; c(A) = A \cap c(B)$\\[.1cm]   
    \textbf{\textsf{PI}} [\emph{path independence}]: &
      $c(A \cup B) = c(c(A) \cup c(B))$.
\end{tabular}\\

Axiom $(\alpha)$ was studied by Chernoff~\cite{Che54}, whereas axioms $(\gamma)$ and $(\beta)$ are due to Sen~\cite{Sen71}.
Axiom $(\rho)$ has been recently introduced in~\cite{CanGiaGreWat16}, in connection to the transitive structure of the relation of revealed preference.
\textsf{WARP} is the most known property of choice consistency, and was introduced by Samuelson~\cite{Sam38} in his original form. 
\textsf{PI} is a property due to Plott~\cite{Plott73}, and its semantics is based on a \textit{divide et impera} principle. 

Upon reformulating these properties also in terms of items -- rather than only using menus --, their semantics becomes clear.
Chernoff's axiom $(\alpha)$ states that any item selected from a menu $B$ must still be selected from any submenu $A \subseteq B$ containing it. 
Sen's axiom $(\gamma)$ says that any item selected from two menus $A$ and $B$ is also selected from the menu $A \cup B$ (if feasible).
The expansion axiom $(\beta)$ can be equivalently written as follows: if $A \subseteq B$, $x,y \in c(A)$ and $y \in c(B)$, then $x \in c(B)$. 
In this form, $(\beta)$ says that if two items are selected from a menu $A$, then they are simultaneously either selected or rejected in any larger menu $B$.
Axiom $(\rho)$ can be equivalently written as follows: if $y \in c(B) \setminus c(B \cup \{x\})$, then $x \in c(B \cup \{x\})$. 
In this form, $(\rho)$ says that if an item $y$ is selected from a menu $B$ but not from the larger menu $B \cup \{x\}$, then the new item $x$ is selected from $B \cup \{x\}$.
\textsf{WARP} summarizes features of contraction and expansion consistency in a single -- and rather strong, despite its name -- axiom, in fact it is equivalent to the conjunction of $(\alpha)$ and $(\beta)$~\cite{Sen71}. 
\textsf{PI} says that an item is selected from a menu $A \cup B$ if and only if it is selected by splitting $A \cup B$ in two submenus, separately choosing from them, and finally selecting from the chosen items: this process is independent of the ``path'', that is, how the elements of $A \cup B$ are split
 for the two preliminary selections.

\section{Liftings}\label{SECT:liftings}

In this section we examine the ``lifting problem": this corresponds to finding necessary and sufficient conditions such that a partial choice satisfying some axioms of consistency can be extended to a total choice satisfying the same axioms. We shall exploit such conditions in the decision results to be presented in Section~\ref{SECT:satProb}.
The next definition makes the notion of lifting formal.
\newcommand\restrict[1]{\raisebox{-.5ex}{$|$}_{#1}}

\begin{definition} \rm \label{DEF:lifting}
    Let $c \colon \Omega \rightrightarrows U$ be a choice. 
    Given a nonempty set $\F$ of sentences of second-order monadic logic, we say that $c$ has the $\F$\textit{-lifting property} if there is a total choice $\cPlus \colon \powPlus \rightrightarrows U$ extending $c$ (i.e., $\cPlus \restrict{\Omega} = c$) and satisfying all formulae in $\F$.
    In this case, $\cPlus$ is called an $\F$\textit{-lifting} of $c$.
    (Of course, we are interested in cases such that $\F$ is a family of axioms of choice consistency.) 
    Whenever $\F$ is a single formula, we simplify notation and write, e.g., $(\alpha)$-lifting, $\textsf{WARP}$-lifting, etc.
    Similarly, we say that $c$ has the \textit{rational lifting property} if there is a total choice $\cPlus$ that is rational and extends $c$. 
\end{definition}

Note that whenever $\F$ is a nonempty set of axioms of choice consistency (which are formulae in prenex normal form where all quantifiers are universal), if a choice has the $\F$-lifting property, then it automatically satisfies all axioms in $\F$. 
The same reasoning applies for the rational lifting property, since it is based on the existence of a binary relation of revealed preference that is fully informative of the choice.

On the other hand, it may happen that a partial choice satisfies some axioms in $\F$ but there is no lifting to a total choice satisfying the same axioms.
The next examples exhibit two instances of this kind.
(To simplify notation, we underline all items that are selected within a menu: for instance $\underline{x}\,y$ and $\underline{x}\,y\,\underline{z}$ stand for, respectively, $c(\{x,y\})=\{x\}$ and $c(\{x,y,z\})=\{x,z\}$. Obviously, we always have $\underline{x}$ for any $\{x\} \in \Omega$, so we can safely omit defining $c$ for singletons.)

\begin{example} \rm \label{EX:no lifting 1}
    Let $U= \{x,y,z\}$ and $\Omega =  \{B \subseteq U : 1 \leq \vert B \vert \leq 2\}$.
    Define a partial choice $c \colon \Omega \rightrightarrows U$ by $\underline{x}\,y$, $\underline{y}\,z$, and $x\,\underline{z}$. 
    This choice is rationalizable by the (cyclic) preference $\precsim$ defined by $x \prec y \prec z \prec x$. 
    However, $c$ does not admit any rational lifting to a total choice $\cPlus$, since we would have $\cPlus(U) = \max (U,\precsim) = \emptyset$.
\end{example}

\begin{example} \rm \label{EX:no lifting 2}
    Let $U= \{x,y,z,w\}$ and $\Omega = \powPlus \setminus \{\{x,w\},\{y,z\},\{y,w\},\{z,w\},U\}$.
    Define a partial choice $c \colon \Omega \rightrightarrows U$ by
    $
    \underline{x}\,\underline{y}\,,\; \underline{x}\,\underline{z}\,,\; x\,\underline{y}\,\underline{z}\,,\;
    \underline{x}\,y\,\underline{w}\,,\; \underline{x}\,\underline{z}\,w\,,\; \underline{y}\,z\,\underline{w}\,.
    $ 
    One can easily check that $c$ satisfies axiom $(\alpha)$ (but it fails to be rationalizable). 
    On the other hand, $c$ admits no $(\alpha)$-lifting, since $\cPlus(U) \neq \emptyset$ violates axiom $(\alpha)$ for any choice $\cPlus$ extending $c$ to the full menu $U$.  
\end{example}

\subsection{Lifting of axiom $(\alpha)$} \label{SECT:lifting alpha}

In this section we characterize the choices that are $(\alpha)$-liftable.
To that end, it is convenient to reformulate axiom $(\alpha)$ in terms of the monotonicity of the rejection map.
We need the following preliminary result, whose simple proof is omitted, and a technical definition.

\begin{lemma}\label{lemmaEquiv}
    Let $A \subseteq B \subseteq U$. 
    For any pair of sets $A',B' \subseteq U$, we have $\:A \cap B' \subseteq A' \;\Longleftrightarrow \; A \setminus A' \subseteq B \setminus B'\,.$
\end{lemma}

\begin{definition} \rm \label{DEF:relativization}      
    Let $c \colon \Omega \rightrightarrows U$ be a choice. 
    For any menu $A \in \powPlus$, let \begin{equation}\label{relativizedDomain}
        \Omega_{A} \defAs \{B \in \Omega : B \subseteq A\}\,.
    \end{equation}
    A set $\mathcal{B} \subseteq \Omega$ of menus is \emph{$\subseteq$-closed w.r.t.\ $\Omega$} if $B \in \mathcal{B}$, for every $B \in \Omega$ such that $B \subseteq \bigcup \mathcal{B}$.   
    \eod
\end{definition}

In view of Lemma~\ref{lemmaEquiv}, axiom $(\alpha)$ can be equivalently rewritten as follows: \begin{equation}\label{equivAlpha}
A \subseteq B \quad \Longrightarrow \quad \oc(A) \subseteq \oc(B)\,.
\end{equation}
In this form, axiom $(\alpha)$ just asserts that enlarging the set of alternatives may only cause the set of neglected members to grow. 
As announced, we have:

\begin{theorem} \label{THM:lifting alpha}
A partial choice $c \colon \Omega \rightrightarrows U$ has the $(\alpha)$-lifting property if and only if the following two conditions hold: 
\begin{enumerate}[label=\text{(\alph*)}, start=1]

\item\label{b} $A \subseteq B \;\; \Longrightarrow \;\; A \cap c(B) \subseteq c(A)$,~ for all $A,B \in \Omega\,$;

\item\label{c} $\bigcup \mathcal{B} \setminus \bigcup_{B \in \mathcal{B}} \overline{c}(B) \neq \emptyset$, for every $\emptyset \neq \mathcal{B} \subseteq \Omega$ such that $\mathcal{B}$ is $\subseteq$-closed w.r.t.\ $\Omega\,$.
\end{enumerate}
\end{theorem}

\begin{proof} 
For necessity, assume that $c \colon \Omega \rightrightarrows U$ can be extended to a total choice $\cPlus$ on $U$ satisfying axiom $(\alpha)$, and let $\oc$ be the associated rejection map of $c$.
Since $c = \cPlus\restrict{\Omega}$, condition~\ref{b} follows immediately from axiom $(\alpha)$ for $\cPlus$. 
To prove that $c$ satisfies condition~\ref{c} as well, let $\mathcal{B}$ be a nonempty $\subseteq$-closed subset of $\Omega$.
By the equivalent formulation (\ref{equivAlpha}) of axiom $(\alpha)$, we obtain $\ocPrime(B) \subseteq \ocPrime(\bigcup \mathcal{B})$ for every $B \in \mathcal{B}$, where $\ocPrime$ is the associated rejection map of $\oc$. Hence
\[  \bigcup_{B \in \mathcal{B}} \ocPrime(B) \; \subseteq \; \ocPrime\big(\bigcup \mathcal{B}\big) \; = \; \bigcup \mathcal{B} \setminus \cPlus\big(\bigcup \mathcal{B}\big)
\]
holds. 
It follows that
\[  \emptyset \; \neq \; \cPlus\big(\bigcup \mathcal{B}\big) \; \subseteq \; \bigcup \mathcal{B} \setminus \bigcup\nolimits_{B \in \mathcal{B}} \ocPrime(B) \; = \; \bigcup \mathcal{B} \setminus \bigcup\nolimits_{B \in \mathcal{B}} \oc(B)\,,
\]
thus showing that \ref{c} holds.
This completes the proof of necessity.

\smallskip

For sufficiency, assume that \ref{b} and \ref{c} hold for the choice $c \colon \Omega \rightrightarrows U$.
For each $A \in \powPlus$, define
\[    \cPlus(A) \: \defAs \: A \setminus \bigcup\nolimits_{B \in \Omega_{A}} \oc(B)\,,
\]
where $\Omega_{C} = \{A \in \Omega : A \subseteq C\}$ (cf.\ Definition~\ref{DEF:relativization}).
In what follows we prove that the map $\cPlus \colon \powPlus \to \pow(U)$ is a well-defined choice, which extends $c$ and satisfies axiom $(\alpha)$.

Since the map $\cPlus$ is obviously contractive by definition, to prove that it is a well-defined choice it suffices to show that it is never empty-valued.
Toward a contradiction, assume that $\cPlus(A) = \emptyset$ for some $A \in \powPlus$.
The definition of $\cPlus$ readily yields
$
  A \; = \; \bigcup\nolimits_{B \in \Omega_{A}} \oc(B) \; \subseteq \; \bigcup \Omega_{A} \; \subseteq \; A\,,
$
which implies $\Omega_{A} \neq \emptyset$ and $\bigcup_{B \in \Omega_{A}} \oc(B) = \bigcup \Omega_{A}$, since $\Omega_{A}$ is $\subseteq$-closed w.r.t.\ $\Omega$.
However, this contradicts \ref{c}. 
Thus, $\cPlus \colon \powPlus \rightrightarrows U$ is a well-defined (total) choice. 
 
Next, we show that $\cPlus$ extends $c$. 
Let $B \in \Omega$. 
Plainly, we have $B \in \Omega_{B}$ and $A \subseteq B$ for each $A \in \Omega_{B}$.
Property \ref{b} yields $A \cap c(B) \subseteq c(A)$, and so $\oc(A) \subseteq \oc(B)$ by Lemma~\ref{lemmaEquiv}. 
Thus, we obtain 
\[\cPlus(B) \: = \: B \setminus \bigcup\nolimits_{A \in \Omega_{B}} \oc(A) \: = \: B \setminus \oc(B) \: = \: c(B)\,,
\]
which proves the claim.

Finally, we show that $\cPlus$ satisfies $(\alpha)$. 
Let $\emptyset \neq B \subseteq C \subseteq U$. 
Since $B \subseteq C$ and $\Omega_{B} \subseteq \Omega_{C}$, we have:\\
\[
B \cap \cPlus(C)  = \; B \cap \left(C \setminus \bigcup\nolimits_{A \in \Omega_{C}} \oc(A)\right) = \; B \setminus \bigcup\nolimits_{A \in \Omega_{C}} \oc(A)
\subseteq \; B \setminus \bigcup\nolimits_{A \in \Omega_{B}}
= \; \cPlus(B)\,.
\]
This proves that axiom $(\alpha)$ holds for $\cPlus$, and the proof is complete.  
\end{proof}

\subsection{Lifting of axiom $(\beta)$} \label{SECT:lifting beta}

Here we prove that any partial choice satisfying axiom $(\beta)$ can be always lifted to a total choice still satisfying axiom $(\beta)$. 
To that end, we need the notion of the \emph{intersection graph} associated to a family of sets $\mathcal{S}$: this is the undirected graph whose nodes are the sets belonging to $\mathcal{S}$, and whose edges are the pairs of distinct intersecting sets $B,B' \in \mathcal{S}$ (i.e., such that $B \cap B' \neq \emptyset$).

\begin{theorem}\label{THM:lifting beta}
A partial choice has the $(\beta)$-lifting property if and only it satisfies axiom $(\beta)$.
\end{theorem}

\begin{proof}
Clearly, axiom $(\beta)$ holds for any choice that admits an extension to a total choice satisfying $(\beta)$.
Thus, it suffices to prove that any choice $c \colon \Omega \rightrightarrows U$ satisfying $(\beta)$ has the $(\beta)$-lifting property. 
For every $A \in \powPlus$, pick an element $u_{A} \in A$, subject only to the condition that $u_{A} \in c(A)$ whenever $A \in \Omega$. 
If $u_{A} \in \bigcup c[\Omega_{A}]$ (where $\Omega_{A} \defAs \{ B \in \Omega : B \subseteq A\}$), then let $\mathcal{C}_{A} \subseteq c[\Omega_{A}]$ be the connected component of the intersection graph associated to the family $c[\Omega_{A}]$ such that $u_{A} \in \bigcup \mathcal{C}_{A}$. 
Then, for $A \in \powPlus$, set
\[
\cPlus(A) \defAs \begin{cases}
\bigcup \mathcal{C}_{A} & \text{if } u_{A} \in \bigcup c[\Omega_{A}]\\
\{u_{A}\} & \text{otherwise}.
\end{cases}
\]
By definition, $\cPlus$ is a total contractive map on $\powPlus$ that is never empty-valued. 
In addition, if $A \in \Omega$, then $\cPlus(A) = \bigcup \mathcal{C}_{A} = c(A)$.
It follows that $\cPlus$ is a well-defined total choice that extends $c$. 

To complete the proof, we only need to show that $\cPlus$ satisfies $(\beta)$. 
Let $D,E \in \powPlus$ be such that $D \subseteq E$ and $\cPlus(D) \cap \cPlus(E) \neq \emptyset$. 
If $|\cPlus(D)| = 1$, then plainly $\cPlus(D) \subseteq \cPlus(E)$. 
On the other hand, if $|\cPlus(D)| > 1$, then $\cPlus(D) = \mathcal{C}_{D}$, hence $\cPlus(E) = \mathcal{C}_{E}$ and $\mathcal{C}_{D} \subseteq \mathcal{C}_{E}$. 
Thus, we obtain again $\cPlus(D) \subseteq \cPlus(E)$, as claimed. \end{proof}

\newcommand{\env}[2]{\mathsf{env}_{#1}(#2)}

\subsection{Lifting of \textsf{WARP}} \label{SECT:lifting of WARP}

Next, we characterize choices that have the \textsf{WARP}-lifting property.
This characterization will be obtained in terms of the existence of a suitable Noetherian total preorder on the collection of the Euler's regions of the union of the choice domain with its image under the given choice. 
(Recall that a \emph{preorder} is a binary relation that is reflexive and transitive. Further, a relation $R$ on $X$ is \emph{Noetherian} if the converse relation $R^{-1}$ is well-founded, i.e., if every nonempty subset of $X$ has an $R$-maximal element.)

Thus, let $c \colon \Omega \rightrightarrows U$ be a partial choice. 
Denote by $\mathcal{E}$ the Euler's diagram of the family 
$
\Omega^{^{+}} \defAs \Omega \cup c[\Omega]\,,
$
namely the partition \\
\centerline{$
\mathcal{E} \defAs \left\{\bigcap \Gamma \:\setminus\: \bigcup (\Omega^{^{+}} \setminus \Gamma) : \emptyset \neq \Gamma \subseteq \Omega^{^{+}} \right\} \setminus \{\emptyset\}
$}
of $\bigcup \Omega^{^{+}}$ formed by all the nonempty sets of the form $\bigcap \Gamma \:\setminus\: \bigcup (\Omega^{^{+}} \setminus \Gamma)$, for $\emptyset \neq \Gamma \subseteq \Omega^{^{+}}$.
Further, for each $A \in \powPlus$, denote by $\env{\mathcal{E}}{A}$ the \emph{envelope of $A$ in $\mathcal{E}$}, namely the collection of regions in $\mathcal{E}$ intersecting $A$; formally, 
$
\env{\mathcal{E}}{A} \defAs \{E \in \mathcal{E} : E \cap A \neq \emptyset\}\,.
$
Observe that, for each $B \in \Omega$, we have $\env{\mathcal{E}}{B} = \{E \in \mathcal{E} : E \subseteq B\}$.
It turns out that the choice $c$ can be lifted to a total choice satisfying \textsf{WARP} if and only if there exists a suitable Noetherian total preorder $\lesssim$ on $\mathcal{E}$ such that
$$
(\exists B \in \Omega) \;\;E \subseteq B  \: \wedge \: E' \subseteq c(B)  \quad  \Longrightarrow \quad E \lesssim E'.
$$
More precisely, we have:

\begin{theorem} \label{THM:lifting WARP}
A partial choice $c \colon \Omega \rightrightarrows U$ has the \textsf{WARP}-lifting property if and only if there exists a total Noetherian preorder $\lesssim$ on the collection $\mathcal{E}$ of Euler's regions of $\Omega \cup c[\Omega]$ such that, for all $B \in \Omega$ and $E,E' \in \mathcal{E}$, the following conditions hold:
\begin{enumerate}[label=(\alph*)]
\item\label{aTheorem3} if $E \subseteq B$ and $E' \subseteq c(B)$, then $E \lesssim E'$;

\item\label{bTheorem3} if 
$E$ is $\lesssim$-maximal in $\env{\mathcal{E}}{B}$, then $E \subseteq c(B)$.
\end{enumerate}
\end{theorem}

\begin{proof}
\textit{(Necessity)} Assume that $c$ can be extended to a total choice $\cPlus$ on $U$ satisfying \textsf{WARP}. We shall show that there exists a total Noetherian preorder $\lesssim$ on $\mathcal{E}$ satisfying conditions \ref{aTheorem3} and \ref{bTheorem3} of the theorem.

For  $E,E' \in \mathcal{E}$, set
\begin{equation}\label{defLesssim}
E \lesssim E' \qquad  \overset{\text{\it\tiny Def}}{\Longleftrightarrow} \qquad \cPlus(E\cup E') \cap E' \neq \emptyset\,.
\end{equation}
In what follows we show that (i) $\lesssim$ is a total preorder, (ii) $\lesssim$ is Noetherian, (iii) $\lesssim$ satisfies condition \ref{aTheorem3}, and (iv) $\lesssim$ satisfies condition~\ref{bTheorem3}. 

\medskip

(i) To prove that $\lesssim$ is a total preorder, observe that $\lesssim$ is reflexive and total by construction.
For transitivity, let $E,E',E'' \in \mathcal{E}$ be such that $E \lesssim E'$ and $E' \lesssim E''$ hold, i.e.,
\begin{align}
\cPlus(E \cup E') \cap E' &\neq \emptyset \label{first_eq}\\
\cPlus(E' \cup E'') \cap E'' &\neq \emptyset \,.\label{second_eq}
\end{align}
We need to show that $\cPlus(E \cup E'') \cap E'' \neq \emptyset$ holds, too.
Plainly, either $\cPlus(E \cup E' \cup E'') \cap E'' \neq \emptyset$, or $\cPlus(E \cup E' \cup E'') \cap (E \cup E') \neq \emptyset$ holds. In the former case, we have
\begin{align*}
\emptyset &\neq \cPlus(E \cup E' \cup E'') \cap E''\\
          &= \cPlus(E \cup E' \cup E'') \cap (E \cup E'') \cap E'' \\
          &= \cPlus(E \cup E'') \cap E'' & \text{[by \textsf{WARP}]}\,.
\end{align*}
In the latter case, \textsf{WARP} and inequality (\ref{first_eq}) yield
\[
\cPlus(E \cup E' \cup E'') \cap (E \cup E') \cap E' = \cPlus(E \cup E') \cap E' \neq \emptyset\,,
\]
so that
\begin{equation}\label{tempDeriv}
\cPlus(E \cup E' \cup E'') \cap E' \neq \emptyset\,.
\end{equation}
Thus, using \textsf{WARP} and the two inequalities (\ref{tempDeriv}) and (\ref{second_eq}), we obtain
\[
\cPlus(E \cup E' \cup E'') \cap (E' \cup E'') \cap E'' = \cPlus(E' \cup E'') \cap E'' \neq \emptyset\,,
\]
hence $\cPlus(E \cup E' \cup E'') \cap E'' \neq \emptyset$, again concluding that $\cPlus(E \cup E'') \cap E'' \neq \emptyset$. 
Thus, in any case, $E \lesssim E''$ holds, as claimed.

\smallskip

(ii) To prove that $\lesssim$ is Noetherian, let $\emptyset \neq \mathcal{A} \subseteq \mathcal{E}$. 
Since $\emptyset \neq \cPlus(\bigcup \mathcal{A}) \subseteq \bigcup \mathcal{A}$, there exists an $\overline{E} \in \mathcal{A}$ such that $\cPlus(\bigcup \mathcal{A}) \cap \overline{E} \neq \emptyset$. 
Thus, it is enough to prove that $E \lesssim \overline{E}$ for each $E \in \mathcal{A}$. 
Let $E \in \mathcal{E}$. 
We have
\begin{align*}
\emptyset &\neq \cPlus\left(\bigcup \mathcal{A}\right) \cap \overline{E}\\
          &= \cPlus\left(\bigcup \mathcal{A}\right) \cap (E \cup \overline{E}) \cap \overline{E}\\
          &= \cPlus(E \cup \overline{E}) \cap \overline{E} & \text{[by \textsf{WARP}]\,,}
\end{align*}
i.e., $E \lesssim \overline{E}$.
This shows that $\overline{E}$ is a maximum of $\mathcal{A}$, proving that the total preorder $\lesssim$ is Noetherian.

\smallskip

(iii) Concerning  \ref{aTheorem3}, let $E \subseteq A$ and $E' \subseteq c(A)$. Hence, $E' \subseteq \cPlus(A)$.
Two applications of \textsf{WARP} yield 
\[
\emptyset \neq \cPlus(E') = \cPlus(A) \cap E' = \cPlus(A) \cap (E \cup E') \cap E' = \cPlus(E \cup E') \cap E'\,,
\]
and so $E \lesssim E'$ holds, proving that \ref{aTheorem3} is satisfied.

\smallskip

(iv) Finally, we prove that also condition \ref{bTheorem3} is satisfied. 
Let $A \in \Omega$, and let $E$ be $\lesssim$-maximal in $\env{\mathcal{E}}{A}$. 
We need to show that $E \subseteq c(A)$. 
Let $E' \subseteq c(A)$. 
By the $\lesssim$-maximality of $E$, we have $E' \lesssim E$, i.e., by (\ref{defLesssim}), $\cPlus(E \cup E') \cap E \neq \emptyset$. 
By \textsf{WARP},
\[
c(A) \cap E = \cPlus(A) \cap E = \cPlus(A) \cap (E \cup E') \cap E = \cPlus(E \cup E') \cap E \neq \emptyset\,,
\]
which plainly yields $E \subseteq c(A)$.
This completes the proof of necessity.

\bigskip

\noindent \textit{(Sufficiency)}
Let $c \colon \Omega \rightrightarrows U$ be a partial choice and assume that there exist a total Noetherian preorder $\lesssim$ on the collection $\mathcal{E}$ of the Euler's regions of $\Omega \cup c[\Omega]$ such that conditions \ref{aTheorem3} and \ref{bTheorem3} hold. 

For every $B \in \powPlus$, set
\begin{equation}\label{defCorrespondence}
\textstyle
\cPlus(B) \defAs \begin{cases}
B \setminus \bigcup \mathcal{E} & \text{if } B \setminus \bigcup \mathcal{E} \neq \emptyset\\
\bigcup \displaystyle\max_{\lesssim} \big(\env{\mathcal{E}}{B}\big) \cap B & \text{otherwise,}
\end{cases}
\end{equation}
where, for any nonempty $\mathcal{A} \subseteq \mathcal{E}$, $\max_{\lesssim} \mathcal{A}$ stands for the collection of the $\lesssim$-maximal members of $\mathcal{A}$. 
To complete the proof, we shall show that (i) $\cPlus$ is a well-defined total choice on $U$, (ii) $\cPlus$ extends $c$, and (iii) $\cPlus$ satisfies \textsf{WARP}. 

\medskip

(i) By definition, $\cPlus$ is a contraction on $\powPlus$. 
Thus, to prove the claim, it suffices to show that $\cPlus(B) \neq \emptyset$, for every $B \in \powPlus$. 
Let $B \in \powPlus$.
If $B \nsubseteq \bigcup \mathcal{E}$, then the result holds trivially. 
Otherwise, let $\emptyset \neq B \subseteq \bigcup \mathcal{E}$. 
Then we have $\cPlus(B) = \bigcup \max_{\lesssim}\big(\env{\mathcal{E}}{B}\big)  \cap B$ and $\env{\mathcal{E}}{B} \neq \emptyset$. 
Since $\lesssim$ is Noetherian, we obtain $\max_{\lesssim} \mathcal{E}_{B} \neq \emptyset$.
Since all members of $\max_{\lesssim}\big(\env{\mathcal{E}}{B}\big)$ intersect $B$, so does their union, proving that $\cPlus(B) \neq \emptyset$ holds in all cases. 

\smallskip

(ii) To show that $\cPlus$ extends $c$, let $A \in \Omega$. 
Since $A \subseteq \bigcup \mathcal{E}$, we have $\cPlus(A) = \bigcup \max_{\lesssim} \big(\env{\mathcal{E}}{A}\big) \cap A$. In fact, since $A \in \Omega \cup c[\Omega]$, then $\cPlus(A) = \bigcup \max_{\lesssim} \big(\env{\mathcal{E}}{A}\big)$. Thus, in order to prove that $\cPlus(A) = c(A)$, we have to show that $\bigcup \max_{\lesssim} \big(\env{\mathcal{E}}{A}\big) = c(A)$. It is then enough to prove that, for $E \in \mathcal{E}$,
\begin{equation}\label{conditionMax}
E \subseteq c(A) \quad \Longleftrightarrow \quad E \in \max_{\lesssim} \big(\env{\mathcal{E}}{A}\big)\,.
\end{equation}
Let $E \in \mathcal{E}$ be such that $E \subseteq c(A)$ and let $E' \in \env{\mathcal{E}}{A}$, so that $E' \subseteq A$ and $E' \in \mathcal{E}$. Thus, by \ref{aTheorem3}, $E' \lesssim E$. The arbitrarity of $E'$ yields that $E \in \max_{\lesssim} \big(\env{\mathcal{E}}{A}\big)$, proving the right-implication of (\ref{conditionMax}). The converse implication follows at once by condition \ref{bTheorem3}.

\smallskip

(iii) Finally, we show that $\cPlus$ satisfies \textsf{WARP}. 
Let $A,B \in \Omega$ be such that $A \subseteq B$ and $A \cap c(B) \neq \emptyset$. 
To prove the claim, we need to show that the equality 
\begin{equation}\label{goal}
\cPlus(A) = A \cap \cPlus(B)
\end{equation}
holds.
If $A \setminus \bigcup \mathcal{E} \neq \emptyset$, then $B \setminus \bigcup \mathcal{E} \neq \emptyset$. 
Hence, by (\ref{defCorrespondence}), equation (\ref{goal}) becomes 
\[
\textstyle
A \setminus \bigcup \mathcal{E} = A \cap (B \setminus \bigcup \mathcal{E})\,,
\]
which plainly holds, since $A \subseteq B$.
On the other hand, suppose $A \subseteq \bigcup \mathcal{E}$.
It follows that $B \subseteq \bigcup \mathcal{E}$, since otherwise (\ref{defCorrespondence}) would yield $\cPlus(B) = B \setminus \bigcup \mathcal{E}$, hence $A \cap \cPlus(B) = \emptyset$, a contradiction. 
Thus, we obtain 
\begin{equation}\label{cAcB}
\cPlus(A) = \textstyle\bigcup \displaystyle\max_{\lesssim} \big(\env{\mathcal{E}}{A}\big) \cap A
\qquad \text{and} \qquad
\cPlus(B) = \textstyle\bigcup \displaystyle\max_{\lesssim} \big(\env{\mathcal{E}}{B}\big) \cap B\,,
\end{equation}
so that to prove (\ref{goal}) it suffices to show that 
\begin{equation}\label{lastGoal}
\textstyle\bigcup \displaystyle\max_{\lesssim} \big(\env{\mathcal{E}}{A}\big) \cap A ~=~ 
\textstyle\bigcup \displaystyle\max_{\lesssim} \big(\env{\mathcal{E}}{B}\big) \cap A\,.
\end{equation}
From our hypothesis $A \cap \cPlus(B) \neq \emptyset$ and (\ref{cAcB}), it follows that $ \max_{\lesssim} \big(\env{\mathcal{E}}{B}\big) \cap \env{\mathcal{E}}{A} \neq \emptyset$. 
Let $E \in \max_{\lesssim} \big(\env{\mathcal{E}}{B}\big) \cap \env{\mathcal{E}}{A}$. 
Since $\env{\mathcal{E}}{A} \subseteq \env{\mathcal{E}}{B}$, we have $E \in \max_{\lesssim} \big(\env{\mathcal{E}}{A}\big)$ too. 
But then, for $E' \in \max_{\lesssim} \big(\env{\mathcal{E}}{B}\big) \cap \env{\mathcal{E}}{A}$, we have $E \lesssim E'$, and therefore $E' \in \max_{\lesssim} \big(\env{\mathcal{E}}{A}\big)$, so that $\max_{\lesssim} \big(\env{\mathcal{E}}{B}\big) \cap \env{\mathcal{E}}{A} \subseteq \max_{\lesssim} \big(\env{\mathcal{E}}{A}\big)$. 
In addition, for $E'' \in \max_{\lesssim} \big(\env{\mathcal{E}}{A}\big)$, we have $E \lesssim E''$, and therefore $E'' \in \max_{\lesssim} \big(\env{\mathcal{E}}{B}\big) \cap \env{\mathcal{E}}{A}$, so that $\max_{\lesssim} \big(\env{\mathcal{E}}{A}\big) \subseteq \max_{\lesssim} \big(\env{\mathcal{E}}{B}\big) \cap \env{\mathcal{E}}{A}$. 
The last two set inclusions yield $\max_{\lesssim} \big(\env{\mathcal{E}}{B}\big) \cap \env{\mathcal{E}}{A} = \max_{\lesssim} \big(\env{\mathcal{E}}{A}\big)$, and therefore
\[
\textstyle\bigcup \displaystyle\max_{\lesssim} \big(\env{\mathcal{E}}{B}\big) \cap A
= \textstyle\bigcup (\displaystyle\max_{\lesssim} \big(\env{\mathcal{E}}{B}\big) \cap \env{\mathcal{E}}{A})
= \textstyle\bigcup \displaystyle\max_{\lesssim} \big(\env{\mathcal{E}}{A}\big)
= \textstyle\bigcup \displaystyle\max_{\lesssim} \big(\env{\mathcal{E}}{A}\big) \cap A\,,
\]
thus proving (\ref{lastGoal}), and in turn completing the proof that $\cPlus$ satisfies \textsf{WARP}.  
\end{proof}

\newcommand{\R}{\mathsf{R}}

\subsection{Lifting of axiom $(\rho)$ for finite choices}\label{SECT:lifting rho}
Before characterizing the choice maps over a finite ground set that are $(\rho)$-liftable, we provide an equivalent and useful formulation of $(\rho)$ in terms of rejection maps.

\begin{lemma}  
The following statements are equivalent for a total choice $c \colon \powPlus \rightrightarrows U$:
\begin{itemize}
	\item[(i)] $c$ satisfies axiom $(\rho)$;
	\item[(ii)] for all menus $A,B \in \powPlus$ such that $A \subseteq B$, we have
	\begin{equation}
		B \setminus A \subseteq \oc(B) \quad\Longrightarrow\quad A \cap \oc(B) \subseteq \oc(A). \label{rhoalter}
	\end{equation}
\end{itemize}
\end{lemma}

\begin{proof} 
	(i)\:$\Longrightarrow$\:(ii). 
	Suppose $c$ satisfies axiom $(\rho)$. 
	Let $A, B \in \powPlus$ be such that $A \subseteq B$ and $B \setminus A \subseteq \oc(B)$, hence $c(B) \subseteq A$.
	The result is trivial if $A = B$.
	Thus assume $A \subsetneq B$, hence $B \setminus A \in \powPlus$. 
	\smallskip
	
	\textsc{Claim:} $c(A) \subseteq c(B)$.  Toward a contradiction, suppose $c(A) \setminus c(B) \neq \emptyset$.  
	An application of axiom $(\rho)$ for the menus $A , B \setminus A$ yields $(B \setminus A) \cap c(B) \neq \emptyset$, which contradicts $c(B) \subseteq A$. 
	\smallskip
	
	Now the Claim readily implies $A \cap \oc(B) \subseteq \oc(A)$, as wanted. 
	\medskip
	
	(ii)\:$\Longrightarrow$\:(i). 
	Suppose (ii) holds for $c$. 
	Let $A, B \in \powPlus$ be such that $c(A) \setminus c(A \cup B) \neq \emptyset$. 
	Toward a contradiction, assume $B \cap c(A \cup B) = \emptyset$. 
	Apply \eqref{rhoalter} to the menus $A$ and $A \cup B$.
	Since the antecedent holds, so does the consequent, hence we get $A \cap \oc(A \cup B) \subseteq \oc(A)$. 
	However, the latter inclusion implies that $c(A) \setminus c(A \cup B) = \emptyset$, which is impossible. 
\end{proof}

Next we give some preparatory definitions and results.

Throughout this section $c \colon \Omega \rightrightarrows U$ will denote a partial choice map over a \emph{finite} ground set $U$. We define by recursion a sequence of maps $\R_{c}^{n} \colon \powPlus \rightarrow \pow(U)$, with $n \geqslant 0$. 
For $A \in \powPlus$, we put: 
\begin{align}
	\R_{c}^{0}(A) &\coloneqq  
	\begin{cases}
		\oc(A) &\text{if } A \in \Omega\\
		\emptyset &\text{otherwise}
	\end{cases}
\label{recursivestart}\\
\R_{c}^{n+1}(A) &\coloneqq \bigcup \left\{ \R_{c}^{n}(B) : A \subseteq B \subseteq U \wedge B \setminus A \subseteq \R_{c}^{n}(B) \right\}, \quad \text{ for } n \geqslant 0.\label{recursivec}
\end{align}

By observing that the recursive equation \eqref{recursivec} can be rewritten as
\[
\R_{c}^{n+1}(A) \coloneqq \R_{c}^{n}(A) \cup \bigcup \left\{ \R_{c}^{n}(B) : A \subsetneq B \subseteq U \wedge B \setminus A \subseteq \R_{c}^{n}(B) \right\},
\]
it follows immediately that the sequence $\{\R_{c}^{n}(A)\}_{n \geqslant 0}$ is monotonic non-decreasing for each $A \in \powPlus$ (hence convergent to $\bigcup_{n \geqslant 0} \R_{c}^{n}(A)$). Thus, we plainly have:
\begin{equation}\label{R0 subsetet Rn}
\R_{c}^{0}(A) \subseteq \R_{c}^{n}(A),
\end{equation}
for all $n \geqslant 0$ and $A \in \powPlus$.

By exploiting the finiteness of the ground set $U$, we can prove the following result.

\begin{lemma}\label{lemma:RncA}
For all $A \in \powPlus$ and $n \geqslant |U| - |A|$, we have
\begin{equation}\label{eq:RncA}
\R_{c}^{n}(A) = \bigcup \big\{ \R_{c}^{|U| - |B|}(B) : A \subseteq B \subseteq U \wedge B \setminus A \subseteq \R_{c}^{|U| - |B|}(B) \big\}.
\end{equation}
Hence, in particular,
\begin{equation}\label{eq:RncA2}
\R_{c}^{m}(A) = \R_{c}^{|U|-|A|}(A)
\end{equation}
holds, for all $m \geqslant |U|$.
\end{lemma}
\begin{proof}
We prove \eqref{eq:RncA} by induction on $|U| - |A|$ (and $n \geqslant |U| - |A|$).

For the base case $|U| - |A| = 0$, we have $A = U$. By iterating \eqref{recursivec} until possible (or, equivalently, by a secondary induction on $n \geqslant 0$), we get
\[
\R_{c}^{n}(U) = \R_{c}^{n-1}(U) = \cdots = \R_{c}^{0}(U),
\]
and since
\[
\bigcup \big\{ \R_{c}^{|U| - |B|}(B) : U \subseteq B \subseteq U \wedge B \setminus A \subseteq \R_{c}^{|U| - |B|}(B) \big\} = \R_{c}^{0}(U),
\]
equation \eqref{eq:RncA} readily follows for $A=U$, namely for $|U| - |A| = 0$.

For the inductive step, let us now assume that $|U| - |A| > 0$, and let $n \geq |U| - |A|$. Again by \eqref{recursivec} and by inductive hypothesis, we have
\begin{align}
\R_{c}^{n}(A) &= \bigcup \big\{ \R_{c}^{n-1}(B) : A \subseteq B \subseteq U \wedge B \setminus A \subseteq \R_{c}^{n-1}(B) \big\} \notag\\
&= \R_{c}^{n-1}(A) \cup \bigcup \big\{ \R_{c}^{|U| - |B|}(B) : A \subsetneq B \subseteq U \wedge B \setminus A \subseteq \R_{c}^{|U| - |B|}(B) \big\}. \label{second}
\end{align}
By iterating \eqref{second} until possible (or, equivalently, by a secondary induction on $n \geqslant |U| - |A|$), we get
\begin{align*}
\R_{c}^{n}(A) &= \R_{c}^{n-1}(A) \cup \bigcup \big\{ \R_{c}^{|U| - |B|}(B) : A \subsetneq B \subseteq U \wedge B \setminus A \subseteq \R_{c}^{|U| - |B|}(B) \big\}\\
&= \R_{c}^{n-2}(A) \cup \bigcup \big\{ \R_{c}^{|U| - |B|}(B) : A \subsetneq B \subseteq U \wedge B \setminus A \subseteq \R_{c}^{|U| - |B|}(B) \big\}\\
& ~~\vdots\\
& = \R_{c}^{|U| - |A|}(A) \cup \bigcup \big\{ \R_{c}^{|U| - |B|}(B) : A \subsetneq B \subseteq U \wedge B \setminus A \subseteq \R_{c}^{|U| - |B|}(B) \big\}\\
& = \bigcup \big\{ \R_{c}^{|U| - |B|}(B) : A \subseteq B \subseteq U \wedge B \setminus A \subseteq \R_{c}^{|U| - |B|}(B) \big\},
\end{align*}
completing the induction proof of \eqref{eq:RncA}.

Equation \eqref{eq:RncA2} follows easily from \eqref{eq:RncA} since, when $m \geqslant |U|$, \emph{a fortiori} we have $m \geqslant |U| - |A|$ and therefore a double application of \eqref{eq:RncA} with $m$ and $|U| - |A|$ yields:
\[
\R_{c}^{m}(A) = \bigcup \big\{ \R_{c}^{|U| - |B|}(B) : A \subseteq B \subseteq U \wedge B \setminus A \subseteq \R_{c}^{|U| - |B|}(B) \big\} = \R_{c}^{|U| - |A|}(A).
\]
\end{proof}

Let us denote by $\R_{c}$ the map $\R_{c}^{|U|}$. The map $\R_{c}$ enjoys the following elementary property.

\begin{lemma}\label{propertyRc}
For all $A,B \in \powPlus$, if $A \subseteq B$ and $B \setminus A \subseteq \R_{c}(B)$, then $\R_{c}(B) \subseteq \R_{c}(A)$.
\end{lemma}

\begin{proof}
It is enough to observe that, by Lemma~\ref{lemma:RncA}, we have
\begin{align*}
\R_{c}(A) &= \R_{c}^{|U|-|A|}(A)\\
&= \bigcup \big\{ \R_{c}^{|U| - |W|}(W) : A \subseteq W \subseteq U \wedge W \setminus A \subseteq \R_{c}^{|U| - |W|}(W) \big\}\\
&= \bigcup \big\{ \R_{c}(W) : A \subseteq W \subseteq U \wedge W \setminus A \subseteq \R_{c}(W) \big\},
\end{align*}
for every $A \in \powPlus$ .
\end{proof}

We are now ready to state and prove the characterization of $(\rho)$-liftability we were after.

\begin{theorem}\label{THM:lifting rho}
A finite choice correspondence $\choice \colon \Omega \rightrightarrows U$ has the $(\rho)$-lifting property if and only if
\begin{enumerate}[label=(\roman*)]
\item $c(A) \cap \R_{c}(A) = \emptyset$, for all $A \in \Omega$; \label{rholifting1}
\item $B \setminus \R_{c}(B) \neq \emptyset$, for all $B \in \powPlus$. \label{rholifting2}
\end{enumerate}
\end{theorem}
\begin{proof}
\emph{(Sufficiency).} Let assume that both \ref{rholifting1} and \ref{rholifting2} are satisfied, and define the map $\cPlus \colon \powPlus \rightarrow \powPlus$ by putting
\[
\cPlus(B) \coloneqq B \setminus \R_{c}(B), \qquad \text{for all } B \in \powPlus.
\]
Plainly, $\cPlus$ is a total choice map.

Next we show that $\cPlus$ agrees with the partial choice $c$ on its domain $\Omega$. Thus, let $A \in \Omega$. Since, by \ref{rholifting1}, $c(A) \cap \R_{c}(A) = \emptyset$, we have $c(A) \subseteq A \setminus \R_{c}(A) = \cPlus(A)$. From \eqref{recursivestart} and \eqref{R0 subsetet Rn}, we have
\[
A \setminus c(A) = \oc(A) = \R_{c}^{0}(A) \subseteq \R_{c}(A).
\]
Hence,
\[
\cPlus(A) = A \setminus \R_{c}(A) \subseteq A \setminus (A \setminus c(A)) = c(A),
\]
and therefore the latter inclusion together with the inclusion $c(A) \subseteq \cPlus(A)$ established earlier implies $\cPlus(A) = c(A)$. The arbitrariness of $A \in \Omega$ yields that $\cPlus$ is an extension of $c$.

It only remains to show that $\cPlus$ satisfies the axiom $(\rho)$, namely (using contrapositive form) that
\begin{equation}\label{contrapositive rho}
B \cap \cPlus(A \cup B) = \emptyset \quad\Longrightarrow\quad \cPlus(A) \subseteq \cPlus(A \cup B)
\end{equation}
holds for all $A,B \in \powPlus$. Thus, let $A,B \in \powPlus$ be such that $B \cap \cPlus(A \cup B) = \emptyset$, that is $B \cap \big((A \cup B) \setminus \R_{c}(A \cup B)\big) = \emptyset$. Hence, 
\begin{align}
&(A \cup B) \setminus A \:\subseteq\: B \:\subseteq\: \R_{c}(A \cup B) \label{firstConsequence}\\
\intertext{and}
&(A \cup B) \setminus \R_{c}(A \cup B) = A \setminus \R_{c}(A \cup B).\label{secondConsequence}
\end{align}
From \eqref{firstConsequence} and Lemma~\ref{propertyRc}, we have $\R_{c}(A \cup B) \subseteq \R_{c}(A)$, and therefore
\[
\cPlus(A) = A \setminus \R_{c}(A) \subseteq A \setminus \R_{c}(A \cup B) = (A \cup B) \setminus \R_{c}(A \cup B) = \cPlus(A \cup B),
\]
proving \eqref{contrapositive rho}. Thus $\cPlus$ is a total choice extending $c$ and complying with axiom $(\rho)$.

\vspace{1cm}

\emph{(Necessity).} Let us now assume that $c$ is $(\rho)$-liftable, and let $\cPlus \colon \powPlus \rightarrow \powPlus$ be a $(\rho)$-lifting of $c$.

\medskip

\noindent \textsc{Claim:} $B \cap \R_{c}(B) \subseteq \ocPlus(B)$, for all $B \in \powPlus$.

\begin{proof}[Proof of the claim]
Preliminarily, we prove by induction on $i \geqslant 0$ the inclusion:
\begin{equation}\label{claim:inclusion}
B \cap \R_{c}^{i}(B) \subseteq \ocPlus(B),
\end{equation}
for all $B \in \powPlus$.

For the base case $i = 0$, by \eqref{recursivestart}, if $B \in \Omega$ then $\R_{c}^{0}(B) \subseteq \oc(B) = \ocPlus(B)$, whereas if $B \in \powPlus \setminus \Omega$ then $\R_{c}^{0}(B) = \emptyset$. In any case, we have $B \cap \R_{c}^{0}(B) \subseteq \ocPlus(B)$.

Next, for the inductive step, let $i > 0$ and let $B \in \powPlus$. Towards a proof of \eqref{claim:inclusion}, let $x \in B \cap \R_{c}^{i}(B)$. From \eqref{recursivec}, there must exist a menu $W \in \powPlus(U)$ such that $B \subseteq W$, $W \setminus B \subseteq \R_{c}^{i-1}(W)$, and $x \in \R_{c}^{i-1}(W)$, so that $x \in W$. Since $W \setminus B \subseteq W \cap \R_{c}^{i-1}(W)$ and, by inductive hypothesis, $W \cap \R_{c}^{i-1}(W) \subseteq \ocPlus(W)$, we have $x \in \ocPlus(W)$ and $W \setminus B \subseteq \ocPlus(W)$. From formulation \eqref{rhoalter} of the axiom $(\rho)$, the latter inclusion yields $B \cap \ocPlus(W) \subseteq \ocPlus(B)$, and therefore we have $x \in \ocPlus(B)$. The arbitrariness of $x \in B \cap \R_{c}^{i}(B)$ yields $B \cap \R_{c}^{i}(B) \subseteq \ocPlus(B)$, completing the proof by induction of \eqref{claim:inclusion}.

From \eqref{claim:inclusion}, it follows immediately
\[
B \cap \R_{c}(B) \subseteq B \cap \R_{c}^{|U|}(B) \subseteq \ocPlus(B),
\]
proving the claim.
\end{proof}

We are now ready to verify the necessity of \ref{rholifting1} and \ref{rholifting2}. Concerning \ref{rholifting1}, let $A \in \Omega$. Then the Claim readily implies 
\[
c(A) \cap \R_{c}(A) \subseteq \ocPlus(A) = \oc(A) = A \setminus c(A),
\]
and therefore we have $c(A) \cap \R_{c}(A) = \emptyset$, proving \ref{rholifting1}.

As for \ref{rholifting2}, let $B \in \powPlus$. Again by the Claim, we have $B \cap \R_{c} \subseteq \ocPlus(B)$, and therefore
\[
\emptyset \neq \cPlus(B) = B \setminus \ocPlus(B) \subseteq B \setminus \R_{c}(B),
\]
proving \ref{rholifting2}, and in turn ending our proof.
\end{proof}

\subsection{Lifting of axiom $(\gamma)$}\label{SECT:lifting gamma}

\newcommand{\mI}[1]{\mathcal{I}_{#1}}
\newcommand{\quarrels}[1]{\mathfrak{Q}^{#1}_{c}}
\newcommand{\mS}[1]{\mathcal{S}\left(#1\right)}
\newcommand{\mN}[1]{\mathcal{N}_{\sol,c}(#1)}
\newcommand{\mNPlus}[1]{\mathcal{N}_{\solPlus,c}(#1)}
\newcommand{\mNmap}{\mathcal{N}_{\sol,c}}
\newcommand{\mNSq}[1]{\mathcal{N}_{\sol,c}[#1]}
\newcommand{\mNstar}[1]{\mathcal{N}_{\sol^{\star}\!\!,c}(#1)}
\newcommand{\mNstarSq}[1]{\mathcal{N}_{\sol^{\star}\!\!,c}[#1]}
\newcommand{\mNprime}[1]{\mathcal{N}_{\sol',c}(#1)}
\newcommand{\mNone}[1]{\mathcal{N}_{\sol_{1},c}(#1)}
\newcommand{\Ic}[1]{\mathcal{I}_{c}(#1)}
\newcommand{\IcPlus}[1]{\mathcal{I}_{\cPlus}(#1)}
\newcommand{\IcSq}[1]{\mathcal{I}_{c}[#1]}
\newcommand{\solPlus}{\sol^{\hbox{\tiny{+}}}}

To characterize those choices that have the $(\gamma)$-lifting property, we first provide some useful definitions and prove some related properties.

\medskip

\newcommand{\mA}{\mathcal{A}}
\newcommand{\mB}{\mathcal{B}}
\newcommand{\mC}{\mathcal{C}}
\newcommand{\mP}{\mathcal{P}}
\newcommand{\mQ}{\mathcal{Q}}
\newcommand{\mR}{\mathcal{R}}
\newcommand{\mT}{\mathcal{T}}
\newcommand{\sol}{\mathfrak{S}}

Given a choice correspondence $c \colon \Omega \rightrightarrows U$, we define the map $\mathcal{I}_{c} \colon \powPlus \rightarrow {\Mi\pow(U)}$ by setting
\[
\Ic{B} \coloneqq \medcup\{ \medcap c[\mB] : \mB \subseteq \Omega,\ \medcup \mB = B\},
\]
for every menu $B \in \powPlus$.

When the choice $c$ satisfies the axiom $(\gamma)$, the following properties hold for the map $\mathcal{I}_{c}$.

\begin{lemma}\label{Ilemma}
Let $c \colon \Omega \rightrightarrows U$ be a choice correspondence that satisfies the axiom $(\gamma)$. Then, for all $B, C \in \powPlus$ we  have
\begin{enumerate}[label=(\alph*)]
\item\label{IlemmaA} $\Ic{B} \cap \Ic{C} \subseteq \Ic{B \cup C}$.
\end{enumerate}

\noindent In particular, if $B \in \Omega$ then

\begin{enumerate}[label=(\alph*),resume]
\item\label{IlemmaB} $\Ic{B} = c(B)$,~~ for every $B \in \Omega$.
\end{enumerate}
\end{lemma}
\begin{proof}
\ref{IlemmaA}~~ Given $B, C \in \powPlus$, then for every $e \in \Ic{B} \cap \Ic{C}$ there exist $\mB, \mC \subseteq \Omega$ such that 
\[
\medcup \mB = B\,,~~ \medcup \mC = C\,,~~ \text{and}~~ 
e \in \medcap c[\mB] \cap \medcap c[\mC]\,.
\] 
Hence $e \in \Ic{B \cup C}$, since $\bigcup(\mB \cup \mC) = B \cup C$ and $e \in \bigcap c[\mB \cup \mC]$. Thus $\Ic{B} \cap \Ic{C} \subseteq \Ic{B \cup C}$.

\smallskip

\ref{IlemmaB}~~ Let us now assume that $B \in \Omega$. Thus $c(B) = 
\medcap c[\{B\}] \subseteq \Ic{B}$, since we plainly have $\{B\} \subseteq \Omega$ and $\bigcup\{B\} = B$.

For the converse inclusion, let $e \in \Ic{B}$. Then there exists a subset $\mB$ of $\Omega$ such that $\medcup \mB = B$
and $e \in \medcap c[\mB]$. Thus, by $(\gamma)$, $e \in c(\bigcup \mB) = c(B)$. Hence, $\Ic{B} \subseteq c(B)$, and in turn $\Ic{B} = c(B)$ holds.
\end{proof}

\begin{definition}[Quarrels]\label{quarrels}
For a choice correspondence $c \colon \Omega \rightrightarrows U$ and an item $e \in U$, we say that a nonempty collection $\mQ \subseteq \powPlus$ is a \emph{$c$-quarrel for $e$} (or just a \emph{quarrel for $e$}, when the choice correspondence is understood), if:
\begin{enumerate}[label=(Q$_{\arabic*}$)]
\item\label{quarrel1} $\medcup \mQ \in \Omega$; and 
\item\label{quarrel2} $e \in \medcap \mQ \cap \oc(\medcup \mQ)$.
\end{enumerate}
We denote by $\quarrels{e}$ the family of all the $c$-quarrels for $e$.
\end{definition}

\begin{remark}\label{firstRemark}
It is immediate to check that, for every choice correspondence $c \colon \Omega \rightrightarrows U$ and menu $A \in \Omega$, the set $\{A\}$ is a quarrel for each item in $\oc(A)$.
\end{remark}

\begin{definition}[Solutions for a choice]\label{quarrelsolution}
Given a choice correspondence $c \colon \Omega \rightrightarrows U$, a map $\sol \colon \powpowPlus \times U \longrightarrow \powpowPlus$ is a \emph{solution for $c$}, if for all items $e \in U$ and quarrels $\mQ \in \quarrels{e}$ for $e$ we have:
\begin{enumerate}[label=(S$_{\arabic*}$)]
\item \label{solution1} $\sol(\mQ,e) \subseteq \mQ$;
\item \label{solution2} $e \notin \bigcup \IcSq{\sol(\mQ,e)}$,~ namely 
$e \notin \Ic{B}$ for every $B \in \sol(\mQ,e)$.
\end{enumerate}
For each solution $\sol$ for $c$, we define a map $\mNmap \colon \powPlus \rightarrow \pow(U)$ by putting
\[
\mN{B} \coloneqq \{ e \in U : B \in \sol(\mQ,e) \text{ for some } \mQ \in \quarrels{e}\}
\]
for each set $B \in \powPlus$. 
\end{definition}

\begin{lemma}\label{Nlemma}
If $\sol$ is a solution for a choice $c \colon \Omega \rightrightarrows U$, then $c(A) = A \setminus \mN{A}$ for every $A \in \Omega$.
\end{lemma}
\begin{proof}
Let $A \in \Omega$. For each item $e \in \oc(A)$ we have $\{A\} \in \quarrels{e}$ (by Remark~\ref{firstRemark}) and $A \in \sol(\{A\},e)$ (by \ref{solution1} and since $\sol(\{A\},e) \neq \emptyset$). Hence $e \in \mN{A}$, and therefore $\oc(A) \subseteq \mN{A}$, so that $A \setminus  \mN{A} \subseteq c(A)$ holds.

To complete the proof, it is enough to establish the converse inclusion $c(A) \subseteq A \setminus \mN{A}$, namely $A \cap \mN{A} = A \setminus (A \setminus \mN{A}) \subseteq \oc(A)$. Thus, let $e \in A \cap \mN{A}$,  then there exists a $\mQ \in \quarrels{e}$ such that $A \in \sol(\mQ,e)$. Hence, by \ref{solution2} and Lemma~\ref{Ilemma}\ref{IlemmaB}, $e \notin \Ic{A} = c(A)$, and therefore $e \in A \setminus c(A) = \oc(A)$, proving that $A \cap \mN{A} \subseteq \oc(A)$ holds, and in turn completing the proof of the lemma.
\end{proof}

\begin{lemma}\label{gammared}
If $\sol$ is a solution for a choice $c \colon \Omega \rightrightarrows U$, then for all $e \in U$ and $\mQ \in \quarrels{e}$ there exists a set $D_{e,\mQ} \in \mQ$ such that the following implication holds 
\begin{equation}\label{gammaredeq}
\medcup \mA = D_{e,\mQ} \:\:\Longrightarrow\:\: e \notin \medcap \mA \setminus \medcup \mNSq{\mA}
\end{equation}
for every $\mA \subseteq \powPlus$.

{\Mi In particular, by setting $\mA \defAs \{D_{e,\mQ}\}$, we have $\medcup \mA = D_{e,\mQ}$ and, by \eqref{gammaredeq}, $e \notin \medcap \mA \setminus \medcup \mNSq{\mA}$.}
\end{lemma}
\begin{proof}
Assume the lemma false. Then there exist $\overline{e} \in U$, $\overline{\mQ} \in \quarrels{\overline{e}}$, and a map $D \mapsto \mA_{D}$ from $\overline{\mQ}$ into $\powpowPlus$ such that
\begin{equation}\label{absurd}
\medcup \mA_{D} = D \quad \text{and} \quad \overline{e} \in \medcap \mA_{D} \setminus \medcup \mNSq{\mA_{D}},
\end{equation}
for every $D \in \overline{\mQ}$.

Let us put
\[
\omQprime \defAs \medcup \big\{ \mA_{D} : D \in \overline{\mQ}\big\}.
\]
Then $\omQprime$ is a quarrel for $\overline{e}$. Indeed,
\[\medcup \omQprime = \medcup \medcup \big\{ \mA_{D} : D \in \overline{\mQ}\big\}
= \medcup  \big\{ \medcup \mA_{D} : D \in \overline{\mQ}\big\}
= \medcup  \big\{ D : D \in \overline{\mQ}\big\}
= \medcup \overline{\mQ},
\]where the second to last equality follows from \eqref{absurd}. By recalling that $\medcup \overline{\mQ} \in \Omega$ and $\overline{e} \in \oc(\medcup \overline{\mQ})$, we readily have $\medcup \omQprime \in \Omega$ (proving \ref{quarrel1}) and $\overline{e} \in \oc(\medcup \omQprime)$. Thus, to prove that also \ref{quarrel2} holds for $\omQprime$, we only need to show that $\overline{e} \in \medcap \omQprime$. Since
\[
\medcap \omQprime = \bigcap_{D \in \omQ} \medcap \mA_{D} = \medcap \big\{ \medcap \mA_{D} : D \in \omQ \big\},
\]
by \eqref{absurd} we have $\overline{e} \in \medcap \omQprime$, hence we have $\omQprime \in \quarrels{\overline{e}}$.

Next, let $A \in \omQprime$. Then $A \in \mA_{D}$ for some $D \in \omQ$, and so by \eqref{absurd} $\overline{e} \notin \mN{\mA_{D}}$. The latter relationship implies that $A \notin \sol(\mQ,\overline{e})$ for all $\mQ \in \quarrels{\overline{e}}$ and, in particular, $A \notin \sol(\omQprime,\overline{e})$. Thus, $\omQprime \cap \sol(\omQprime,\overline{e}) = \emptyset$. Since by \ref{solution1} $\sol(\omQprime,\overline{e}) \subseteq \omQprime$, we have $\sol(\omQprime,\overline{e}) = \emptyset$, which is a contradiction. Hence, the lemma follows.
\end{proof}

\begin{lemma}\label{gammasufflemma}
For every solution $\sol$ for a given choice $c \colon \Omega \rightrightarrows U$, there exists another solution $\sol^{\star}$ for $c$ such that 
\begin{equation}\label{gammasufflemmaA}
\mNstar{A} \subseteq \mN{A}
\end{equation}
and
\begin{equation}\label{gammasuffeq}
\big(e \in \cap \mA \ \wedge\ e \notin \medcup \mNstarSq{A} \big)\quad \Longrightarrow \quad e \notin \mNstar{\medcup \mA} 
\end{equation}
hold for all $A \in \powPlus$, $\mA \in \powpowPlus$, and $e \in U$.
\end{lemma}
\begin{proof}
If $\sol$ satisfies condition \eqref{gammasuffeq}, we are done, as $\sol$ trivially satisfies \eqref{gammasufflemmaA}. Thus, let us assume that $\sol$ does not satisfy \eqref{gammasuffeq}.

For each $e \in U$ and $\mQ \in \quarrels{e}$, we define the following collections of menus
\newcommand{\mPQe}{\mQ^{1}_{e}}
\newcommand{\mPQAe}{(\mQ_{A})^{1}_{e}}
\newcommand{\mPQeBar}{\overline{\mQ}^{\raisebox{-2.2pt}{\scriptsize$1$}}_{e}}
\newcommand{\mPQeABar}{(\overline{\mQ_{A}})^{\raisebox{-2.2pt}{\scriptsize$1$}}_{e}}
\newcommand{\mRQe}{\mQ^{2}_{e}}
\newcommand{\mTQe}{\mQ_{e}}
\newcommand{\mQtilde}{\widetilde{\mQ}}
\newcommand{\mQtildeOne}{\widetilde{\mQ}_{e}^{\raisebox{0pt}{\scriptsize$1$}}}
\begin{align*}
\mPQe &\coloneqq \big\{ D \in \mQ : e \notin \medcap \mathcal{A} \setminus \medcup \mNSq{\mA}, \text{~for all~ $\mA \subseteq \powPlus$ such that $\medcup \mA = D \big\}$}, \\
\mRQe &\coloneqq \medcup \big\{ \mA \subseteq \powPlus : \medcup \mA \in \mQ  \ \wedge \ e \in \medcap \mA \setminus \medcup \mNSq{\mA} \big\},\\
\mTQe &\coloneqq \mPQe \cup \mRQe,
\end{align*}
and prove the following facts:
\begin{enumerate}[label=(\Alph*)]
\item\label{fact1} $\mTQe \neq \emptyset$,

\item\label{fact2} $\medcup \mTQe = \medcup \mQ$, and

\item\label{fact3} $\mTQe$ is a quarrel for $e$.
\end{enumerate}

Concerning \ref{fact1}, letting $D_{e,\mQ}$ be any set that verifies condition \eqref{gammaredeq} of Lemma~\ref{gammared} (whose existence is stated in the same lemma), we plainly have $D_{e,\mQ} \in \mPQe$. Hence, the collection $\mPQe$, and \emph{a fortiori} $\mTQe$, is nonempty.

As for \ref{fact2}, we plainly have $\medcup \mPQe \subseteq \medcup \mQ$ and $\medcup \mRQe \subseteq \medcup \mQ$, so that $\medcup \mTQe \subseteq \medcup \mQ$. Concerning the reverse implication $\medcup \mQ \subseteq \medcup \mTQe$, for every $x \in \medcup \mQ$ we have $x \in D_{x}$ for some $D_{x} \in \mQ$. If $e \notin \medcap \mA \setminus \medcup \mNSq{\mA}$ for all $\mA \subseteq \powPlus$ such that $\medcup \mA = D_{x}$, then $D_{x} \in \mPQe$ and so $D_{x} \subseteq \medcup \mPQe$; else, letting $\mA_{x} \subseteq \powPlus$ be such that $\medcup \mA_{x} = D_{x}$ and $e \in \medcap \mA_{x} \setminus \medcup \mNSq{\mA_{x}}$, we have $\mA_{x} \subseteq \mRQe$ and so $D_{x} = \medcup \mA_{x} \subseteq \medcup \mRQe$. Thus, in any case, $x \in D_{x} \subseteq \medcup \mPQe \cup \medcup \mRQe = \medcup \mTQe$, proving $\medcup \mQ \subseteq \medcup \mTQe$ by the arbitrariness of $x \in \medcup \mQ$.

Finally, concerning \ref{fact3}, by recalling that $\mQ$ is a quarrel for $e$, we have $\medcup \mQ \in \Omega$ and $e \in \medcap \mQ \cap \oc(\medcup \mQ)$. Then, in view of \ref{fact2}, we have $\medcup \mTQe \in \Omega$ (hence, condition \ref{quarrel2} of Definition~\ref{quarrels} is satisfied by $\mTQe$) and $e \in \oc(\medcup \mTQe)$. Thus, to show that $\mTQe$ is a quarrel for $e$, it only remains to check that $e \in \medcap \mTQe$, namely that $e \in A$ for every $A \in \mTQe$. Let then $A \in \mTQe$. If $A \in \mPQe$, then $A \in \mQ$ and  since $e \in \medcap \mQ$ we plainly have $e \in A$. On the other hand, if $A \in \mRQe$, then $A \in \mA$ for some $\mA$ such that $e \in \medcap \mA$, and so we have $e \in A$ again. Thus, $e \in \medcap \mTQe$, as desired, and therefore $\mTQe$ is a quarrel for $e$.

\medskip

Next, we define a map $\sol'$ over $\powpowPlus \times U$ by putting, for all $e \in U$ and $\mQ \in \quarrels{e}$,
\[
\sol'(\mQ,e) \coloneqq \sol(\mTQe,e),
\]
and prove the following claims related to $\sol'$:

\begin{claim}\label{Claim1}
$\emptyset \neq \sol'(\mQ,e) \subseteq \mPQe$, for all $e \in U$ and $\mQ \in \quarrels{e}$.
\end{claim}

\begin{claim}\label{Claim2}
$\sol'$ is a solution for $c$.
\end{claim}

\begin{claim}\label{Claim3}
$\mNprime{A} \subseteq \mN{A}$, for all $A \in \powPlus$.
\end{claim}

\renewcommand\qedsymbol{$\blacksquare$}
\begin{proof}[Proof of Claim~\ref{Claim1}]
Let $e \in U$ and $\mQ \in \quarrels{e}$. To begin with, by the very definition of $\sol'(\mQ,e)$ and $\mTQe$, and since $\sol$ is a solution for $c$, we have
\[
\emptyset \neq \sol'(\mQ,e) = \sol(\mTQe,e) \subseteq \mTQe = \mPQe \cup \mRQe.
\]
Thus, to complete the proof of the claim it suffices to show that we have $\sol(\mTQe,e) \cap \mRQe = \emptyset$. To this purpose, let $A \in \mRQe$. Then $e \notin \mN{A}$ so that $A \notin \sol(\mQ',e)$ for all $\mQ' \in \quarrels{e}$, and in particular $A \notin \sol(\mTQe,e)$. Therefore $\sol(\mTQe,e) \cap \mRQe = \emptyset$ holds, completing the proof of the claim.
\end{proof}

\begin{proof}[Proof of Claim~\ref{Claim2}]
Let $e \in U$ and $\mQ \in \quarrels{e}$. From Claim~\ref{Claim1},
\[
\sol'(\mQ,e) \subseteq \mPQe \subseteq \mQ,
\]
hence $\sol'$ satisfies condition \ref{solution1} of Definition~\ref{quarrelsolution} about solutions. Concerning condition \ref{solution2}, let $A \in \sol'(\mQ,e)$. Thus, $A \in \sol(\mTQe,e)$ and therefore $e \notin \Ic{A}$, proving \ref{solution2} for $\sol'$ by the arbitrariness of $e$, $\mQ$, and $A$. 

Having proved that $\sol'$ satisfies conditions \ref{solution1} and \ref{solution2}, it follows that $\sol'$ is a solution for $c$.
\end{proof}

\begin{proof}[Proof of Claim~\ref{Claim3}]
Let $A \in \powPlus$ and let $e \in \mNprime{A}$. Then there exists $\mQ \in \quarrels{e}$ such that $A \in \sol'(\mQ,e) = \sol(\mTQe,e)$, and therefore we readily have $e \in \mN{A}$. The arbitrariness of $e \in \mNprime{A}$ yields $\mNprime{A} \subseteq \mN{A}$, proving the claim.
\end{proof}

\renewcommand\qedsymbol{$\openbox$}

\bigskip

Back to the proof of Lemma~\ref{gammasufflemma}, we are now ready to define a solution $\sol^{\star}$ for $c$ satisfying conditions \eqref{gammasufflemmaA} and \eqref{gammasuffeq} of the lemma: for all $e \in U$ and $\mQ \in \quarrels{e}$, we put
\[
\sol^{\star}(\mQ,e) = \big(\sol(\mQ,e) \cup \sol'(\mQ,e)\big) \setminus \mPQeBar,
\]
where $\mPQeBar \defAs \mQ \setminus \mPQe$.

Firstly, we show that $\sol^{\star}$ is actually a solution for $c$. Let again $e \in U$ be a generic item and $\mQ \in \quarrels{e}$ a generic quarrel for $e$. Since $\sol$ and $\sol'$ are solutions for $c$ (by assumption and by Claim~\ref{Claim2}, respectively), we have $\sol(\mQ,e) \subseteq \mQ$ and (from Claim~\ref{Claim1} and the very definition of $\mPQe$) $\sol'(\mQ,e) \subseteq \mPQe \subseteq \mQ$. Hence,
\[
\emptyset \neq \sol'(\mQ,e) \subseteq \sol^{\star}(\mQ,e) = \big(\sol(\mQ,e) \cup \sol'(\mQ,e)\big) \setminus \mPQeBar \subseteq \mQ,
\]
proving that $\sol^{*}$ is indeed a map into $\powpowPlus$ satisfying condition \ref{solution1} to be a solution for $c$. Concerning condition \ref{solution2}, let $A \in \sol^{\star}(\mQ,e)$. Then, either $A \in \sol(\mQ,e)$ or $A \in \sol'(\mQ,e)$. In any case, we have $e \notin \Ic{A}$, in view of condition \ref{solution2} for the solutions $\sol$ and $\sol'$. Hence, $\sol^{\star}$ enjoys condition \ref{solution2} too, and so it is a solution for $c$.

Next we check that the solution $\sol^{\star}$ for $c$ satisfies conditions \eqref{gammasufflemmaA} and \eqref{gammasuffeq} of the lemma.

Concerning condition \eqref{gammasufflemmaA}, let $A \in \powPlus$ and $e \in \mNstar{A}$. Then there exists a quarrel $\mQ$ for $e$ such that $A \in \sol^{\star}(\mQ,e)$. If $A \in \sol(\mQ,e)$, then $e \in \mN{A}$. On the other hand, if $A \in \sol'(\mQ,e)$, then $e \in \mNprime{A}$, and so, by Claim~\ref{Claim3}, $e \in \mN{A}$ again. Thus, $\mNstar{A} \subseteq \mN{A}$, proving \eqref{gammasufflemmaA}.

Finally, we prove that $\sol^{\star}$ fulfils condition \eqref{gammasuffeq} too. By way of contradiction, assume that this is not so, and let $e \in U$ and $\mA \in \powpowPlus$ be such that 
\[
e \in A \cap B, \quad e \notin \medcup \mNstarSq{\mA}, \quad \text{and}~~  e \in \mNstar{\medcup \mA}.
\]
In view of the latter membership, there exists a quarrel $\mQ$ for $e$ such that $\medcup \mA \in \sol^{\star}(\mQ,e) \subseteq \mQ$, and so $\medcup \mA \notin \mPQeBar$. Hence, $\medcup \mA \in \mPQe$ and therefore $e \notin \medcap \mA \setminus \medcup \mNSq{\mA}$, which in turn implies $e \in \medcup \mNSq{\mA}$.

We partition $\mA$ into the following subfamilies:
\[
\mA_{1} \defAs \big\{ A \in \mA : e \in \mN{\mA} \big\} \quad \text{and} \quad \mA_{2} \defAs \big\{ A \in \mA : e \notin \mN{\mA} \big\}.
\]
Plainly $\mA_{1} \neq \emptyset$. Let $A$ be any set in $\mA_{1}$, and let $\mQ_{A} \in \quarrels{e}$ be such that $A \in \sol(\mQ_{A}, e) \subseteq \mQ_{A}$. Since $e \notin \mNstar{A}$, we have $A \notin \sol^{\star}(\mQ_{A},e)$ and so $A \in \mPQeABar = \mQ_{A} \setminus \mPQAe$, from which $A \notin \mPQAe$ follows. Thus, there exists $\mB_{A} \subseteq \powPlus$ such that
\begin{equation}\label{condDoubleStar}
\medcup \mB_{A} = A \quad \text{and} \quad e \in \mB_{A} \setminus \medcup \mNSq{\mB_{A}}.
\end{equation}
Notice that, while doing so, we have defined a map $A \mapsto \mB_{A}$ from $\mA_{1}$ into $\powpowPlus$ such that \eqref{condDoubleStar} holds for every $A \in \mA_{1}$.

Next, let us consider the collection $\mB_{1} = \medcup \{\mB_{A} : A \in \mA_{1}\}$. Plainly $e \in \medcap \{\medcap \mB_{A} : A \in \mA_{1}\} =  \medcap \mB_{1}$, and since $e \in \medcap \mA$, we also have $e \in \medcap (\mB_{1} \cup \mA_{2})$. In addition, we have:
\begin{align*}
\medcup (\mB_{1} \cup \mA_{2}) &= \medcup \big( \medcup \{ \mB_{A} : A \in \mA_{1} \} \big) \cup \medcup \mA_{2}\\
&= \medcup \{ \medcup \mB_{A} : A \in \mA_{1} \}  \cup \medcup \mA_{2}\\
&= \medcup \{ A : A \in \mA_{1} \}  \cup \medcup \mA_{2}\\
&= \medcup \mA_{1}  \cup \medcup \mA_{2}\\
&= \medcup \big( \mA_{1}  \cup \mA_{2} \big)\\
&= \medcup \mA,
\end{align*}
and therefore $\medcup (\mB_{1} \cup \mA_{2}) \in \mPQe$, so that $e \notin \medcap (\mB_{1} \cup \mA_{2}) \setminus \medcup \mNSq{\mB_{1} \cup \mA_{2}}$. 

Since $e \in \medcap (\mB_{1} \cup \mA_{2})$, we have 
\begin{equation}\label{lastContradiction}
e \in \medcup \mNSq{\mB_{1} \cup \mA_{2}}.
\end{equation}
On the other hand, from \eqref{condDoubleStar} it follows that $e \notin \medcup \big\{ \medcup \mNSq{\mB_{A}} : A \in \mA_{1} \big\} = \medcup \mNSq{\mB_{1}}$ and $e \notin \medcup \mNSq{\mA_{2}}$, so that $e \notin \medcup \mNSq{\mB_{1} \cup \mA_{2}}$, which contradicts \eqref{lastContradiction}.

Having shown that the assumption that $\sol^{\star}$ does not fulfil condition \eqref{gammasuffeq} led to a contradiction, it follows that $\sol^{\star}$ must indeed satisfy \eqref{gammasuffeq}. We have already proved that $\sol^{\star}$ satisfies also condition \eqref{gammasufflemmaA}, hence we can conlude that $\sol^{*}$ is a solution for $c$ of the type we were seeking for.
\end{proof}

We are now ready to state and prove the announced characterization for $(\gamma)$-lifting. 

\begin{theorem}\label{THM:lifting gamma}
Let $c \colon \Omega \rightrightarrows U$ be a choice correspondence that satisfies $(\gamma)$. Then $c$ has the $(\gamma)$-lifting property 
if and only if there exists a solution $\sol$ for $c$ so that  $A \setminus \mN{A} \neq \emptyset$, for all $A \in \powPlus$.
\end{theorem}
\begin{proof}
\emph{(Necessity).} Let $\cPlus$ be a total choice \emph{extending} $c$ and  satisfying axiom $(\gamma)$, and let $\solPlus$ be any map over $\powpowPlus \times U$ such that
\[
\solPlus(\mQ,e) = \{A \in \mQ : e \in \ocPlus(A)\},
\]
for each $e \in U$ and each $\cPlus$-quarrel $\mQ$ for $e$.

We prove that $\solPlus$ is a solution for $\cPlus$ such that $A \setminus \mN{A} \neq \emptyset$, for all $A \in \powPlus$. Since any $c$-quarrel for $e \in U$ is trivially a $\cPlus$-quarrel for $e$, $\solPlus$ will also be a solution for $c$.

By its very definition, $\solPlus$ satisfies condition \ref{solution1} of solutions. Concerning condition \ref{solution2}, let $e \in U$, $\mQ$ a $\cPlus$-quarrel for $e$, and $A \in \solPlus(\mQ,e)$ (hence, $e \in \ocPlus(A)$ holds), and assume for contradiction that $e \in \IcPlus{A}$. Then there exists $\mA \subseteq \powPlus$ such that $\medcup \mA = A$ and $e \in \medcap \cPlus[\mA]$. By axiom $(\gamma)$, we have $\medcap \cPlus[\mA] \subseteq \cPlus(\medcup \mA) = \cPlus(A)$, and therefore $e \in \cPlus(A)$, contradicting the fact noted above that $e \in \ocPlus(A)$. Thus, $e \notin \IcSq{A}$ must hold. From the arbitrariness of $e$, $\mQ$, and $A$, it therefore follows that also condition \ref{solution2} holds for the map $\solPlus$, proving that $\solPlus$ is indeed a solution for $\cPlus$, and for $c$ too (as already observed).

To complete the necessity part of the proof, we are left with showing that $A \setminus \mNPlus{A} \neq \emptyset$, for all $A \in \powPlus$. Thus, let $A \in \powPlus$, and let $e \in \mNPlus{A}$. Then $A \in \solPlus(\mQ,e)$ for some $\mQ \in \quarrels{e}$, and so $e \in \ocPlus(A)$ (and $A \in \medcup \quarrels{e}$). Hence, $\mNPlus{A} \subseteq \ocPlus(A)$ and so 
\[
\emptyset ~\neq~ \cPlus(A) = A \setminus \ocPlus(A) ~\subseteq~ A \setminus \mNPlus{A}.
\]

\smallskip

\noindent \emph{(Sufficiency).} Let us now assume that there exists a solution $\sol$ for $c$ such that $A \setminus \mN{A} \neq \emptyset$, for all $A \in \powPlus$. By Lemma~\ref{gammasufflemma}, there exists a solution $\sol^{\star}$ for $c$ such that conditions \eqref{gammasufflemmaA} and \eqref{gammasuffeq} hold for $\sol^{\star}$. From \eqref{gammasufflemmaA}, it follow that $\emptyset \neq A \setminus \mN{A} \subseteq A \setminus \mNstar{A}$, for all $A \in \powPlus$. Hence, the map $\cPlus \colon \powPlus \rightrightarrows U$ defined by 
\[
\cPlus(A) \defAs A \setminus \mNstar{A}, \qquad \text{for } A \in \powPlus,
\]
is a choice correspondence. In addition, by Lemma~\ref{Nlemma}, $\cPlus(A) = c(A)$ holds for every $A \in \Omega$, and so $\cPlus$ is an extension of $c$.

Finally, we prove that $\cPlus$ satisfies axiom $(\gamma)$. To this purpose, let $\mA \in \powpowPlus$, and let $e \in \medcap \cPlus[\mA] = \bigcap_{A \in \mA} \big( A \setminus \mNstar{A} \big)$. Then $e \in \medcap \mA$ and $e \notin \medcup \mNstarSq{\mA}$, and so by \eqref{gammasuffeq} we have $e \notin \mNstar{\medcup \mA}$. Hence, $e \in \medcup \mA \setminus \mNstar{\medcup \mA} = \cPlus(\medcup \mA)$, and so $\medcap \cPlus[\mA] \subseteq \cPlus(\medcup \mA)$, proving that $\cPlus$ satisfies axiom $(\gamma)$, by the genericity of $\mA \in \powpowPlus$.
\end{proof}

\bigskip

Besides analyzing the lifting properties for single axioms, it is interesting to study characterizations for partial choices that can be extended to rationalizable total choices or to choices satisfying \textsf{WARP}.

\subsection{Lifting of rationalizability}\label{SEC:lifting razionalizable}

We need some preliminary definitions.

\begin{definition}\label{DEF:rationalizable selection}
Given a choice map $c: \Omega \rightrightarrows U$, not necessarily total, we define the following binary relation $P_{c}$ over $U$
\[
P_{c} \defAs \{(x,y) \in U \times U : x \neq y \wedge (\forall A \in \Omega)\ (x,y \in A \rightarrow y \in \oc(A))\}
\]
and put
\[
P_{c}(y,A) \defAs \{ (x,y) \in P_{c} : x \in A\}
\]
for each $A \in \Omega$ and $y \in \oc(A)$.\\
Finally, we say that a map $\sigma$ with domain $\{(y,A) : A \in \Omega \wedge y \in \oc(A)\}$ is a {\em selection for} $c$ if $\sigma(y,A) \in P_{c}(y,A)$, for all $(y,A)$ in its domain. In addition, we say that $\sigma$ is \emph{acyclic} if so is its range $\{\sigma(y,A) : A \in \Omega \wedge y \in \oc(A)\}$ w.r.t.\ the relation $P_{c}$.
\end{definition}

The following theorem provides a characterization for a partial choice to be liftable to a total rationalizable choice.

\begin{theorem}\label{THM:lifting rationalizable}
A partial choice $c : \Omega \rightrightarrows U$ can be extended to a total rationalizable choice if and only if $c$ has an acyclic selection (w.r.t.\ the relation $P_{c}$).
\end{theorem}
\begin{proof}
\emph{(Necessity).} Suppose $\cPlus$ is a total rationalizable choice that extends $c$, and let $\rightarrowtail$ be the asymmetric\footnote{See Footnote~\ref{foot:asymmetric}.} binary relation that rationalizes it, namely such that
\begin{equation}\label{eq:rationalization}
\cPlus(A) = \{ y \in A : x \rightarrowtail y \text{ for no } x \in A \}, \quad \text{for all $A \in \powPlus$.}
\end{equation}
Then we have
\begin{equation}\label{PcPlus}
P_{\cPlus} = \{(x,y) \in U \times U : x \rightarrowtail y \}.
\end{equation}
Indeed, if $x \rightarrowtail y$ then plainly $x \neq y$, otherwise $\ocPlus(\{x,y\}) = \emptyset$. In addition, if $x,y \in A$ then by \eqref{eq:rationalization} $y \in \ocPlus(A)$, for every $A \in \powPlus$. Conversely, if $x \neq y \wedge (\forall A \in \powPlus)\ (x,y \in A \rightarrow y \in \oc(A))$, then in particular $y \in \ocPlus(\{x,y\})$, and therefore $x \rightarrowtail y$ must hold.

Plainly, 
\begin{equation}\label{PcPlusSubseteq}
P_{\cPlus} \subseteq P_{c}
\end{equation}
and
\begin{equation}\label{PcPlusCap}
P_{\cPlus}(y,A) = P_{c}(y,A) \cap P_{\cPlus}, \quad \text{for all $A \in \Omega$ and $y \in c(A)$.}
\end{equation}

Notice that, by \eqref{eq:rationalization} and \eqref{PcPlus}, $P_{\cPlus}(y,A)$ is nonempty for each $y \in \ocPlus(A)$. Hence, by \eqref{PcPlusCap}, $P_{c}(y,A)$ is nonempty as well, for all $A \in \Omega$ and $y \in \oc(A)$.

We conclude the first half of the proof by observing that any map $\sigma$ with domain $\{(y,A) : A \in \Omega \wedge y \in \oc(A)\}$ and such that $\sigma(y,A) \in P_{\cPlus}(y,A)$, for all $(y,A)$ in its domain, is an acyclic selection for $c$. Indeed, if $x_{0},x_{1},\ldots,x_{n},x_{0}$ formed a $P_{c}$-cycle, then by \eqref{PcPlusSubseteq} they would also form a $P_{\cPlus}$-cycle, and so we would have $\cPlus(\{x_0,x_1,\ldots,x_n\}) = \emptyset$ (as by \eqref{PcPlus} $\cPlus$ is rationalized by the relation $P_{\cPlus}$), contradicting the totality of $\cPlus$.

\emph{(Sufficiency).} Suppose that $\sigma$ is an acyclic selection for $c$. Let $\rightarrowtail$ be the range of $\sigma$ 
and let $\cPlus$ be the total choice generated by $\rightarrowtail$. Since $\rightarrowtail $ is acyclic,
$\cPlus$ is a total choice, which is rationalized by $\rightarrowtail$. It only remains to prove that $\cPlus$ extends the choice $c$. Thus, let $A \in \Omega$. If $y \in \oc(A)$, then letting $(x,y) = \sigma(y,A) \in P_{c}(y,A) \subseteq P_{c}$ we have $x \in A$ and $x \rightarrowtail y$. Thus, $y \in \ocPlus(A)$, and so $\oc(A) \subseteq \ocPlus(A)$. Conversely, if $y \in  \ocPlus(A)$, then there exists an $x \in A$ so that $x \rightarrowtail y$, and therefore $(x,y) = \sigma(y,B)$ for some $B \in \Omega$, and so  $(x,y) \in P_{c}(y,B) \subseteq P_{c}$. Hence, by the very definition of $P_{c}$, we have $y \in \oc(A)$, and so $\ocPlus(A) \subseteq \oc(A)$. Hence, $\ocPlus(A) = \oc(A)$, and therefore we can conclude that $\cPlus(A) \subseteq c(A)$, thus showing that $\cPlus$ is a total rationalizable choice that extends $c$.
\end{proof}

\section{The satisfiability problem in presence of a choice correspondence}\label{SECT:satProb}

We are now ready to define the syntax and semantics of the Boolean set-theoretic language extended by a choice correspondence, denoted by $\BSTC$, of which we shall study the satisfiability problem.

\subsection{Syntax of $\BSTC$}
The language $\BSTC$ involves:
\begin{itemize}
\item denumerable collections $\mathcal{V}_{0}$ and $\mathcal{V}_{1}$ of \emph{individual} and \emph{set} variables, respectively;

\item the constant $\varnothing$ (empty set);

\item operation symbols: $\cdot\cup\cdot$, $\cdot\cap\cdot$, $\cdot\setminus\cdot$ , $\{\cdot\}$, $\choice(\cdot)$ (choice map);

\item predicate symbols: $\cdot=\cdot$, $\cdot\subseteq\cdot$, $\cdot\in\cdot$.
\end{itemize}
\emph{Set terms} of $\BSTC$ are recursively defined as follows:
\begin{itemize}
\item set variables and the constant $\varnothing$ are set terms;

\item if $T, T_{1},T_{2}$ are set terms and $x$ is an individual variable, then
$
T_{1} \cup T_{2},~~ T_{1} \cap T_{2},~~ T_{1} \setminus T_{2},~~ \choice(T),~~ \{x\}
$
are set terms.
\end{itemize}
The \emph{atomic formulae} (or \emph{atoms}) of $\BSTC$ have one of the following two forms
\[
T_{1} = T_{2}, \qquad  T_{1} \subseteq T_{2}, \]
where $T_{1},T_{2}$ are set terms. Atoms and their negations are called \emph{literals}.\\[.1cm]
Finally, \emph{$\BSTC$-formulae} are propositional combinations of $\BSTC$-atoms by means of the usual logical connectives $\land$, $\lor$, $\lnot$, $\implies$, $\iff$.

We regard $\{x_{1},\ldots,x_{k}\}$ as a shorthand for the set term $\{x_{1}\} \cup \cdots \cup \{x_{k}\}$. Likewise, $x \in T$ and $x = y$ are regarded as shorthands for $\{x\} \subseteq T$ and $\{x\} = \{y\}$, respectively.

\emph{Choice terms} are $\BSTC$-terms of type $\choice(T)$, whereas \emph{choice-free terms} are $\BSTC$-terms which do not involve the choice map $\choice$ (at any level of nesting). We refer to $\BSTC$-formulae containing only choice-free terms as $\BSTC^{-}$-formulae.\footnote{Up to minor syntactic differences, $\BSTC^{-}$-formulae are essentially \textsf{2LSS}-formulae, whose decision problem has been solved (see, for instance, \cite[Exercise 10.5]{CanOmoPol01}).}

\subsection{Semantics of $\BSTC$}

We first describe the \emph{unrestricted semantics of $\BSTC$}, when the choice operator is not required to satisfy any particular consistency axiom.

A \emph{set assignment} is a pair $\model=(U,M)$, where $U$ is any nonempty collection of objects, called the \emph{domain} or \emph{universe} of $\model$, and $M$ is an assignment over the variables of $\BSTC$ such that
\begin{itemize}
\item $\M x \in U$, for each individual variable $x \in \mathcal{V}_{0}$;

\item $\M \varnothing \defAs \emptyset$;

\item $\M X \subseteq U$, for each set variable $X \in \mathcal{V}_{1}$;

\item $\M \choice$ is a total choice correspondence over $U$.
\end{itemize}
Then, recursively, we put
\begin{itemize}
\item $\M{(T_{1} \otimes T_{2})} \defAs \M T_{1} \otimes \M T_{2}$, where $T_{1},T_{2}$ are set terms and $\otimes \in \{\cup,\cap,\setminus\}$;

\item $\M{\{x\}} \defAs  \{\M x\}$, where $x$ is an individual variable;

\item $\M {(\choice(T))} \defAs \M\choice(\M T)$, where $T$ is a set term.
\end{itemize}
Satisfiability of any $\BSTC$-formula $\psi$ by $\model$ (written $\model \models \psi$) is defined by putting
\[\begin{array}{rclrcr}
\model &\models& T_{1} \star T_{2}   &\qquad\text{iff}\qquad& \M T_{1} \star \M T_{2}\,,
\end{array}
\]
for $\BSTC$-atoms $T_{1} \star T_{2}$ (where $T_{1},T_{2}$ are set terms and $\star \in \{=,\subseteq\}$), and by interpreting logical connectives according to their classical meaning.

For a $\BSTC$-formula $\psi$,  if $\model \models \psi$ (i.e.,
$\model$ \emph{satisfies} $\psi$), then $\model$ is said to be a
\emph{$\BSTC$-model for $\psi$}.  A $\BSTC$-formula is said to be
\emph{satisfiable} if it has a $\BSTC$-model.  
Two $\BSTC$-formulae $\varphi$ and $\psi$ are \emph{equivalent} if they share exactly the same $\BSTC$-models; they are \emph{equisatisfiable} if one is satisfiable if and only if so is the other (possibly by different $\BSTC$-models). 

The \emph{satisfiability problem} (or \emph{decision problem}) for $\BSTC$ asks for an effective procedure (or \emph{decision procedure}) to establish whether any given $\BSTC$-formula is satisfiable or not.

\smallskip

We shall also address the satisfiability problem for $\BSTC$ under other semantics: specifically, the \emph{$(\alpha)$-semantics}, the \emph{$(\beta)$-semantics}, and the \textsf{WARP}-\emph{semantics} (whose satisfiability relations are denoted by $\models_{\alpha}$, $\models_{\beta}$, and $\models_{\textsf{WARP}}$, respectively). These differ from the unrestricted semantics in that the interpreted choice map $\M \choice$ is required to satisfy axiom $(\alpha)$ in the first case, axiom $(\beta)$ in the second case, and axioms $(\alpha)$ and $(\beta)$ conjunctively (namely \textsf{WARP}) in the latter case.

\smallskip

\subsection{The decision problem for $\BSTC$-formulae}\label{appendixDecProc}

The satisfiability problem for $\BSTC^{-}$- and $\BSTC$-formulae under the various semantics are \NP-hard, as the satisfiability problem for propositional logic can readily be reduced to any of them (in linear time).
In the cases of $(\alpha)$- and \textsf{WARP}-semantics, we shall prove \NP-completeness only under the additional hypothesis that the number of choice terms is constant, otherwise, in both cases, we have to content ourselves with a \NEXP\ complexity. As a by-product, it will follow that the satisfiability problem for $\BSTC^{-}$ is \NP-complete. On the other hand, we shall prove that the satisfiability problem for $\BSTC$-formulae under the unrestricted and the $(\beta)$-semantics can be reduced polynomially to the satisfiability problem for $\BSTC^{-}$-formulae, thereby proving their \NP-completeness.

\smallskip

Let $\varphi$ be a $\BSTC$-formula, $\mathsf{V}_{0} \subseteq \mathcal{V}_{0}$ and $\mathsf{V}_{1} \subseteq \mathcal{V}_{1}$ the collections of individual and set variables occurring in $\varphi$, respectively, and $\terms$ the collection of the set terms occurring in $\varphi$. For convenience, we shall assume that $\varnothing \in \terms$. Let also
\begin{equation}\label{choiceLiterals}
\choice(T_{1}),~\ldots,~\choice(T_{k})
\end{equation}
be the distinct choice terms occurring in $\varphi$, with $k \geqslant 0$ (when $k=0$, $\varphi$ is a $\BSTC^{-}$-formula). 
 
Without loss of generality, we may assume that $\varphi$ is in \emph{choice-flat form}, namely that all the terms $T_{1},\ldots,T_{k}$ in (\ref{choiceLiterals}) are choice-free. In fact, if this were not the case, then, for each choice term $\choice(T)$ in $\varphi$ occurring inside the scope of a choice symbol and such that $T$ is choice-free, we could replace in $\varphi$ all occurrences of $\choice(T)$ by a newly introduced variable $X_{T}$ and add the conjunct $X_{T} = \choice(T)$ to $\varphi$, until no choice term is left which properly contains a choice subterm. It is an easy matter to check that the resulting formula is in choice-flat form, it is equisatisfiable with $\varphi$ (under any of our semantics), and its size is linear in the size of $\varphi$. 

Without disrupting satisfiability, we may add to $\varphi$ the following formulae:
\begin{description}
\item[\emph{choice conditions}:] $\varnothing \neq \choice(T_{i}) \;\land\; \choice(T_{i}) \subseteq T_{i}$, ~~for $i=1,\ldots,k$;

\item[\emph{single-valuedness conditions}:] $T_{i} = T_{j} \implies \choice(T_{i}) = \choice(T_{j})$, ~~for all distinct $i,j = 1,\ldots,k$,
\end{description}
since they are plainly true in any $\BSTC$-assignment.
In this case the size of $\varphi$ could have up to a quadratic increase. However, the total number of terms remains unchanged.\footnote{See Footnote~\ref{FootnoteFiner}.} For the sake of simplicity, we shall assume that $\varphi$ includes its choice and single-valuedness conditions, and thereby say that it is \emph{complete}. 

Notice that the above considerations hold irrespectively of the semantics adopted. 

In the sections which follow, we study the satisfiability problem for complete $\BSTC$-formulae under the various semantics described earlier. We start our course with the \textsf{WARP}-semantics.

\subsubsection{\textsf{WARP}-semantics}\label{WARPsemantics}

We first derive some necessary conditions for $\varphi$ to be satisfiable and later prove their sufficiency. Hence, to begin with, let us assume that $\varphi$ is satisfiable under the \textsf{WARP}-semantics and let $\model = (U,M)$ be a model for it. 
Let $\M{\mathcal{R}_{\varphi}}$ be the Euler's diagram of $\M{\terms} \defAs \{\M T : T \in \terms\}$.
Notice that, for each region $\rho \in \M{\mathcal{R}_{\varphi}}$ and term $T \in \terms$, either $\rho \subseteq \M T$ or $\rho \cap \M T = \emptyset$. Thus, to each $\rho \in \M{\mathcal{R}_{\varphi}}$,  there corresponds a Boolean map $\pi_{\rho} \colon \terms \rightarrow \{\true,\false\}$ over $\terms$ (where we have identified the truth values \textsf{true} and \textsf{false} with $\true$ and $\false$, respectively) such that
\begin{equation}\label{defPlaces}
\pi_{\rho}(T) = \true \iff \rho \subseteq \M T \iff \rho \cap \M T \neq \emptyset\,,~~~~\text{for } T \in \terms\,.
\end{equation}
Let $\M{\Pi_{\varphi}} \defAs \{\pi_{\rho} : \rho \in \M{\mathcal{R}_{\varphi}}\}$. Hence, we have:
\begin{enumerate}[label=\text{(\roman*)}]
\item\label{zeroPlace} $\pi(\varnothing) = \false$, for each $\pi \in \M{\Pi_{\varphi}}$;

\item\label{firstPlace} $\pi(T_{1} \cup T_{2}) = \pi(T_{1}) \lor \pi(T_{2})$, for each map $\pi \in \M{\Pi_{\varphi}}$ and set term $T_{1} \cup T_{2}$ in $\varphi$;

\item\label{secondPlace} $\pi(T_{1} \cap T_{2}) = \pi(T_{1}) \land \pi(T_{2})$, for each map $\pi \in \M{\Pi_{\varphi}}$ and set term $T_{1} \cap T_{2}$ in $\varphi$;

\item\label{thirdPlace} $\pi(T_{1} \setminus T_{2}) = \pi(T_{1}) \land \lnot \pi(T_{2})$, for each map $\pi \in \M{\Pi_{\varphi}}$ and set term $T_{1} \setminus T_{2}$ in $\varphi$.
\end{enumerate}
In addition, we have $\M T = \bigcup \{\rho \in \M{\mathcal{R}_{\varphi}} : \pi_{\rho}(T) = \true\}$, for every $T \in \terms$.

By uniformly replacing the atomic formulae in $\varphi$ with propositional variables, in such a way that different occurrences of the same atomic formula are replaced by the same propositional variable and different atomic formulae are replaced by distinct propositional variables, we can associate to $\varphi$ its \emph{propositional skeleton} $\mathsf{P}_{\varphi}$ (up to variables renaming). For instance, the propositional skeleton of
\begin{equation}\label{prop-example}
((X = Y \setminus X \:\land\: Y = X \cup \choice(X_{1})) \Longrightarrow Z \neq \varnothing) \Longleftrightarrow (X = Y \setminus X \Longrightarrow (Y = X \cup \choice(X_{1})) \Longrightarrow Z = \varnothing))
\end{equation}
is the propositional formula 
\[
((P_{1} \land P_{2}) \Longrightarrow \neg P_{3}) \Longleftrightarrow (P_{1} \Longrightarrow (P_{2} \Longrightarrow P_{3}))\,.
\]
Plainly, a necessary condition for $\varphi$ to be satisfiable (by a $\BSTC$-model) is that its skeleton $\mathsf{P}_{\varphi}$ is propositionally satisfiable (however, the converse does not hold in general). 

\smallskip
A collection $\mathcal{A}$ of atoms of $\varphi$ is said to be \emph{promising} for $\varphi$ if the valuation which maps to \textsf{true} the propositional variables corresponding to the atoms in $\mathcal{A}$ and to \textsf{false} the remaining ones satisfies the propositional skeleton $\mathsf{P}_{\varphi}$. For instance, in the case of (\ref{prop-example}), all collections of its  atoms not containing both $X = Y \setminus X$ and $Y = X \cup \choice(X_{1})$ are promising for (\ref{prop-example}).

Let $\mathcal{A}_{\varphi}^{+}$ be the collection of the atoms in $\varphi$ satisfied by $\model$ and $\mathcal{A}_{\varphi}^{-}$ the collection of the remaining atoms in $\varphi$, namely those that are disproved by $\model$. It can be easily checked that $\mathcal{A}_{\varphi}^{+}$ is promising. In addition, for every atom $T_{1} = T_{2}$ in $\varphi$ and $\pi \in \M{\Pi_{\varphi}}$, we have $\pi(T_{1}) = \pi(T_{2})$ if and only if $T_{1} = T_{2}$ is in $\mathcal{A}_{\varphi}^{+}$. Likewise, for every atom $T_{1} \subseteq T_{2}$ in $\varphi$ and $\pi \in \M{\Pi_{\varphi}}$, we have $\pi_{\rho}(T_{1}) \leqslant \pi_{\rho}(T_{2})$ if and only if $T_{1} \subseteq T_{2}$ is in $\mathcal{A}_{\varphi}^{+}$. Thus, in particular, for every atom $T_{1} = T_{2}$ in $\mathcal{A}_{\varphi}^{-}$, there exists a map $\pi \in \M{\Pi_{\varphi}}$ such that $\pi(T_{1}) \neq \pi(T_{2})$. Likewise, for every atom $T_{1} \subseteq T_{2}$ in $\mathcal{A}_{\varphi}^{-}$, there exists a map $\pi \in \M{\Pi_{\varphi}}$ such that $\pi(T_{1}) > \pi(T_{2})$.

\begin{definition}[Places]\label{places}\rm
Any Boolean map $\pi$ on $\terms$\, for which the above properties \ref{zeroPlace}--\ref{thirdPlace} hold for each set term in $\varphi$ is called a \emph{place for $\varphi$}.

For a given set $\mathcal{A}$ of atoms occurring in $\varphi$, a set $\Pi$ of places for $\varphi$ is $\mathcal{A}$-\emph{ample} if 
\begin{enumerate}[label=\text{(\roman*)},start=5]
\item\label{fourthPlace} $\pi(T_{1}) = \pi(T_{2})$ for each $\pi \in \Pi$, provided that the atom $T_{1} = T_{2}$ is not in $\mathcal{A}$;

\item\label{fifthPlace} $\pi(T_{1}) \leqslant \pi(T_{2})$ for each $\pi \in \Pi$, provided that the atom $T_{1} \subseteq T_{2}$ is not in $\mathcal{A}$;

\item\label{sixthPlace} $\pi(T_{1}) \neq \pi(T_{2})$ for some $\pi \in \Pi$, if the atom $T_{1} = T_{2}$ belongs to $\mathcal{A}$;

\item\label{seventhPlace} $\pi(T_{1}) > \pi(T_{2})$ for some $\pi \in \Pi$, if the atom $T_{1} \subseteq T_{2}$ belongs to $\mathcal{A}$.
\eod
\end{enumerate}
\end{definition}
Notice that conditions \ref{zeroPlace}--\ref{fifthPlace} are universal, whereas conditions \ref{sixthPlace}--\ref{seventhPlace} are existential.

The considerations made just before Definition~\ref{places} yield that $\M{\Pi_{\varphi}}$ is an $\mathcal{A}_{\varphi}^{-}$-ample set of places for $\varphi$. However, in order to later establish some tight complexity results, it is convenient to enforce a polynomial bound for the cardinality of the set of places in terms of the size $|\varphi|$ of $\varphi$ (where, for instance, $|\varphi|$ could be defined as the number of nodes in the syntax tree of $\varphi$). We do this as follows: for each atom $T_{1} = T_{2}$ (resp., $T_{1} \subseteq T_{2}$) in $\mathcal{A}_{\varphi}^{-}$, we select a place $\pi_{\rho}$, with $\rho \in \M{\mathcal{R}_{\varphi}}$ such that $\rho \subseteq \M{T_{1}} \Longleftrightarrow \rho \nsubseteq \M{T_{2}}$ (resp., $\rho \subseteq \M{T_{1}} \setminus \M{T_{2}}$) holds, and call their collection $\Pi_{1}$. Plainly, we have $|\Pi_{1}| \leq |\mathcal{A}_{\varphi}^{-}|$. Notice that $\Pi_{1}$ is $\mathcal{A}_{\varphi}^{-}$-ample.\footnote{\label{FootnoteFiner}A finer construction would yield an $\mathcal{A}_{\varphi}^{-}$-ample set of places $\Pi_{1}'$ such that $|\Pi_{1}'| \leq |\terms|$. Notice that $|\Pi_{1}| = \mathcal{O}(|\Pi_{1}'|^{2})$.}

Conditions \ref{zeroPlace}--\ref{thirdPlace} take care of the structure of set terms in $\varphi$ but those of the form $\{x\}$ or $\choice(T)$, conditions \ref{fourthPlace} and \ref{fifthPlace} take care of the atoms in $\varphi$ deemed to be positive, whereas conditions \ref{sixthPlace} and \ref{seventhPlace} take care of the remaining atoms in $\varphi$, namely those deemed to be negative.

To take care of set terms of the form $\{x\}$ in $\varphi$, we observe that, for every $x \in \mathsf{V}_{0}$, there exists a unique Euler's region $\rho_{x} \in \M{\mathcal{R}_{\varphi}}$ such that $\M x \in \rho_{x}$. Let  $\pi^{x}$ be the place corresponding to $\rho_{x}$ according to (\ref{defPlaces}), namely $\pi^{x} \defAs \pi_{\rho_{x}}$, and put $\Pi_{2} \defAs \{\pi^{x} : x \in \mathsf{V}_{0}\}$. 

\begin{definition}[Places at variables]\label{placeAtVariables}\rm
Let $x$ be an individual variable occurring in $\varphi$. A \emph{place (for $\varphi$) at the variable $x$} is any place $\pi$ for $\varphi$ such that $\pi(\{x\}) = \true$.\eod
\end{definition}

Next, we take care of choice terms. Thus, let $\Omega \defAs \{\M{T_{1}},\ldots,\M{T_{k}}\}$ and let ${\mathcal{E}}$ be the Euler's diagram of $\Omega \cup \M\choice[\Omega]$. Notice that each region in ${\mathcal{E}}$ is a disjoint union of regions in $\M{\mathcal{R}_{\varphi}}$; moreover, the partial choice $\M\choice\restrict{\Omega}$ over the choice domain $\Omega$ enjoys the $\textsf{WARP}$-lifting property. Thus, by Theorem~\ref{THM:lifting WARP}, there exists a total Noetherian preorder $\lesssim$ on ${\mathcal{E}}$ such that, for all $\sigma',\sigma'' \in {\mathcal{E}}$ and $\M{T} \in \Omega$, we have:
\begin{enumerate}[label=(\Alph*)]
\item\label{Aabove}  if $\sigma' \subseteq \M{T}$ and $\sigma'' \subseteq \M\choice(\M{T})$, then $\sigma' \lesssim \sigma''$;

\item\label{Babove}  if $\sigma'$ is $\lesssim$-maximal in $\mathsf{env}_{{\mathcal{E}}}(\M{T})$, then $\sigma' \subseteq \M\choice(\M{T})$.
\end{enumerate}
For each region $\sigma \in \mathcal{E}$, let us select a place $\pi_{\rho}$, such that $\rho \in \M{\mathcal{R}_{\varphi}}$ and $\rho \subseteq \sigma$, and call their collection $\Pi_{3}$. Set $\Pi \defAs \Pi_{1} \cup \Pi_{2} \cup \Pi_{3}$. Plainly, $|\Pi| \leqslant |\mathcal{A}| + |\mathsf{V}_{0}| + 2^{k}$, $\Pi$ is $\mathcal{A}_{\varphi}^{-}$-ample, and $\pi^{x}$ is the sole place in $\Pi$ at the variable $x$, for each individual variable $x$ in $\varphi$. To ease notation, for $\pi \in \Pi$, $\Pi' \subseteq \Pi$, and $T \in \terms$,  we shall also write (i) $\pi \subseteq T$ for $\pi(T) = \true$, (ii) $\Pi' \subseteq T$ for $\pi' \subseteq T$, for every $\pi' \in \Pi'$, (iii) $\Pi' \ni\in T$ for $\pi' \subseteq T$, for some $\pi' \in \Pi'$.
\renewcommand{\P}{\mathcal{P}}
For each $\sigma \in \mathcal{E}$, let $\Pi_{\sigma} \defAs \{\pi_{\rho} : \rho \in \M{\mathcal{R}_{\varphi}} \text{ and } \rho \subseteq \sigma\}$, and call $\P$ their collection. 
Then, by \ref{Aabove} and \ref{Babove} above, there exists a total Noetherian preorder $\precsim$ on $\P$ such that, for all $\Pi',\Pi'' \in \P$ and $i \in \{1,\ldots,k\}$, the following conditions hold:
\begin{enumerate}[label=(\Alph*')]
\item\label{AabovePrime}  if $\Pi' \subseteq T_{i}$ and $\Pi'' \subseteq \choice(T_{i})$, then $\Pi' \precsim \Pi''$;

\item\label{BabovePrime}  if $\Pi'$ is $\precsim$-maximal in $\{\Pi^{*} \in \P : 
\Pi^{*} \ni\in T_{i}\}$, then $\Pi' \subseteq \choice(T_{i})$.
\end{enumerate}

Summing up, we have the following result:
\begin{lemma}\label{necessity2LSS}
Let $\varphi$ be a $\BSTC$-formula in choice-flat form, $\mathsf{V}_{0}$ the set of individual variables occurring in it, and $\choice(T_{1}),~\ldots,~\choice(T_{k})$ the choice terms occurring in it. If $\varphi$ is satisfiable under the \textsf{WARP}-semantics, then there exist an $\mathcal{A}$-ample set $\Pi$ of places for $\varphi$ such that $|\Pi| \leq |\mathcal{A}| + |\mathsf{V}_{0}| + 2^{k}$, for some promising set $\mathcal{A}$ of atoms in $\varphi$, and a map $x \mapsto \pi^{x}$ from $\mathsf{V}_{0}$ into $\Pi$ such that $\pi^{x}$ is the sole place in $\Pi$ at the variable $x$, for $x \in \mathsf{V}_{0}$. In addition, if
$
\Pi_{\choice} \defAs \big\{\pi \in \Pi : \pi \subseteq T_{i} \text{ or } \pi \subseteq \choice(T_{i})\text{, for some } i \in \{1,\ldots,k\} \big\}\,,
$
$\sim_{\choice}$ is the equivalence relation on $\Pi_{\choice}$ such that
\\
\centerline{$
\pi \sim_{\choice} \pi' \quad \Longleftrightarrow \quad \pi(T_{i}) = \pi'(T_{i}) \text{ ~and~ } \pi(\choice(T_{i})) = \pi'(\choice(T_{i}))\text{\,,~ for every } i = 1,\ldots,k\,,
$}
and $\P \defAs \Pi_{\choice}/\sim_{\choice}$, then there exists a total Noetherian preorder $\precsim$ on $\P$ such that conditions \ref{AabovePrime} and \ref{BabovePrime} are satisfied for all $\Pi',\Pi'' \in \P$ and $i \in \{1,\ldots,k\}$.
\eod
\end{lemma}

Next we show that the conditions in the preceding lemma are also sufficient for the satisfiability of our $\BSTC$-formula $\varphi$ under the \textsf{WARP}-semantics. Thus, let $\Pi$, $\mathcal{A}$, $x \mapsto \pi^{x}$, $\Pi_{\choice}$, $\sim_{\choice}$, and $\P$ be such that the  conditions in Lemma~\ref{necessity2LSS} are satisfied. Let $U_{_{\Pi}}$ be any set of cardinality $|\Pi|$, and $\pi \mapsto a_{\pi}$ any injective map from $\Pi$ onto $U_{_{\Pi}}$. We define an interpretation $M_{_{\Pi}}$ over the variables in $\mathsf{V}_{0} \cup \mathsf{V}_{1}$ and the choice terms $\choice(T_{1}),~\ldots,~\choice(T_{k})$ occurring in $\varphi$, as if they were set variables, by putting:
\newcommand{\piM}[1]{#1^{\scriptscriptstyle M_{\Pi}}}
\begin{equation}\label{finalModel}
\begin{array}{rcll}
\piM x &\defAs& a_{\pi^{x}}&\text{for } x \in \mathsf{V}_{0}\\
\piM X &\defAs& \{a_{\pi} \in U_{_{\Pi}} : \pi \in \Pi \land \pi \subseteq X \}~~~~~&\text{for } X \in \mathsf{V}_{1}\\
\piM{(\choice(T_{i}))} &\defAs& \{a_{\pi} \in U_{_{\Pi}} : \pi \in \Pi \land \pi \subseteq \choice(T_{i}) \}~~~~~&\text{for } i =1,\ldots,k\,.
\end{array}
\end{equation}
Notice that the choice map $\choice$ is not interpreted by $M_{_{\Pi}}$. However, it is not hard to check that the set assignment $\model_{_{\Pi}} \defAs (U_{_{\Pi}},M_{_{\Pi}})$ satisfies exactly the atoms in $\mathcal{A}$, provided that choice terms in $\varphi$ are regarded as set variables with no internal structure, rather than as compound terms. Indeed, after putting $\piM \pi \defAs \{a_{\pi}\}$, for $\pi \in \Pi$, we have $\piM T = \bigcup_{\pi \subseteq T} \piM \pi$, for every $T \in \terms$. Thus, if $T_{1} = T_{2}$ does not belong to $\mathcal{A}$, then by condition \ref{fourthPlace} we have $\piM{T_{1}} = \piM{T_{2}}$, whereas if $T_{1} = T_{2}$  belongs to $\mathcal{A}$, then by  \ref{sixthPlace} we have $\piM{T_{1}} \neq \piM{T_{2}}$. Analogously for atoms $T_{1} \subseteq T_{2}$ in $\varphi$.
Hence, by the promisingness of $\mathcal{A}$, $\model_{_{\Pi}}$ satisfies also $\varphi$.

Next we define $M_{_{\Pi}}$ also over the choice map $\choice$ in such a way that $\piM \choice$ is a total choice over $U$ that satisfies \textsf{WARP}, and $\model_{_{\Pi}} \models \varphi$ holds.
Let $\Omega_{_{\Pi}} \defAs \{\piM{T_{1}},\ldots,\piM{T_{k}}\}$. We begin by defining $\piM{\choice}$ over $\Omega_{_{\Pi}}$ in the most natural way, namely by putting $\piM{\choice}(\piM{T_{i}}) \defAs \piM{(\choice(T_{i}))}$, for $i = 1,\ldots,k$. By the choice and the single-valuedness conditions in $\varphi$, $\piM{\choice}$ so defined is a choice over the domain $\Omega_{_{\Pi}}$. In addition, it can be checked that the existence of a total Noetherian preorder on $\P$ satisfying conditions \ref{AabovePrime} and \ref{BabovePrime} yield the existence of a total Noetherian preorder on the collection of the Euler's regions of $\Omega_{_{\Pi}} \cup \piM{\choice}[\Omega_{_{\Pi}}]$ satisfying conditions \ref{aTheorem3} and \ref{bTheorem3} of Theorem~\ref{THM:lifting WARP}. Thus, the latter theorem readily implies that $\piM{\choice}$ has the \textsf{WARP}-lifting property, so that it can be extended to a total choice on $U$ satisfying \textsf{WARP}, proving that $\varphi$ is satisfiable under the \textsf{WARP}-semantics.

Together with Lemma~\ref{necessity2LSS}, the above argument yields the following result:
\begin{theorem}\label{WARPTheorem}
Let $\varphi$ be a complete $\BSTC$-formula in choice-flat form, $\mathsf{V}_{0}$ the set of individual variables occurring in it, and $\choice(T_{1}),~\ldots,~\choice(T_{k})$ the choice terms occurring in it. Then $\varphi$ is satisfiable under the \textsf{WARP}-semantics if and only if there exist 
(i) an $\mathcal{A}$-ample set $\Pi$ of places for $\varphi$ such that $|\Pi| \leq |\mathcal{A}| + |\mathsf{V}_{0}| + 2^{k}$, for some promising set $\mathcal{A}$ of atoms in $\varphi$, (ii) a map $x \mapsto \pi^{x}$ from $\mathsf{V}_{0}$ into $\Pi$, such that $\pi^{x}$ is the sole place in $\Pi$ at the variable $x$, for $x \in \mathsf{V}_{0}$, and (iii) a total Noetherian preorder $\precsim$ on $\P$, 
where $\P \defAs \Pi_{\choice}/\sim_{\choice}$ (with $\Pi_{\choice} \defAs \big\{\pi \in \Pi : \pi \subseteq T_{i} \text{ or } \pi \subseteq \choice(T_{i})\text{, for some } i \in \{1,\ldots,k\} \big\}\,,$ and $\sim_{\choice}$  the equivalence relation on $\Pi_{\choice}$ such that
$\pi \sim_{\choice} \pi'  \Longleftrightarrow  \pi(T_{i}) = \pi'(T_{i}) \text{ ~and~ } \pi(\choice(T_{i})) = \pi'(\choice(T_{i}))\text{\,,~ for } i = 1,\ldots,k$) such that the above conditions \ref{AabovePrime} and \ref{BabovePrime} are satisfied for all $\Pi',\Pi'' \in \P$ and $i \in \{1,\ldots,k\}$.
\eod
\end{theorem}

We have already observed that the satisfiability problem for $\BSTC$-formulae under the \textsf{WARP}-semantics is  \NP-hard. In addition, the previous theorem implies at once that it belongs to \NEXP. However, if we restrict to $\BSTC$-formulae with a constant number of choice terms, the disequality $|\Pi| \leq |\mathcal{A}| + |\mathsf{V}_{0}| + 2^{k}$ in Theorem~\ref{WARPTheorem} yields $|\Pi| = \mathcal{O}(|\mathcal{A}| + |\mathsf{V}_{0}|) =\mathcal{O}(|\varphi|)$, thereby providing the following complexity result:
\begin{theorem}
Under the \textsf{WARP}-semantics, the satisfiability problem for $\BSTC$-formulae with $\mathcal{O}(1)$ distinct choice terms is \NP-complete.\eod
\end{theorem}
\medskip
As a by-product, Theorem~\ref{WARPTheorem} yields a solution to the satisfiability problem for $BSTC^{-}$-formulae. Indeed, in the case of $BSTC^{-}$-formulae, Theorem~\ref{WARPTheorem} becomes:

\begin{theorem}\label{BSTC-Theorem}
Let $\varphi$ be a $\BSTC^{-}$-formula, and $\mathsf{V}_{0}$ the set of individual variables occurring in it. Then $\varphi$ is satisfiable if and only if there exist (i) an $\mathcal{A}$-ample set $\Pi$ of places for $\varphi$ such that $|\Pi| \leq |\mathcal{A}| + |\mathsf{V}_{0}|$, for some promising set $\mathcal{A}$ of atoms in $\varphi$, and (ii) a map $x \mapsto \pi^{x}$ from $\mathsf{V}_{0}$ into $\Pi$, such that $\pi^{x}$ is the sole place in $\Pi$ at the variable $x$, for $x \in \mathsf{V}_{0}$.
\eod
\end{theorem}

Hence, we also have:
\begin{theorem}\label{NPcompl2LSS}
The satisfiability problem for $\BSTC^{-}$-formulae is \NP-complete.\eod
\end{theorem}

\subsubsection{Unrestricted semantics}

As above, let $\varphi$ be a complete $\BSTC$-formula in choice flat-form.
Plainly, if $\varphi$ is satisfiable under unrestricted semantics, so is its \emph{$\BSTC^{-}$-reduction} obtained from $\varphi$ by regarding the choice terms as set variables with no internal structure.

Conversely, let us assume that the $\BSTC^{-}$-reduction $\varphi_{1}$ of $\varphi$ is satisfiable, and let $\model_{1} = (U,M_{1})$ be a model for $\varphi_{1}$. 
We define a total choice correspondence $c \colon \powPlus \rightrightarrows U$ on $U$ by putting, for every $A \in \powPlus$,
\begin{equation}\label{defMapc}
c(A) \defAs \begin{cases}
A & \text{if } A \notin \{ \M T_{1},\ldots, \M T_{k}\}\\
\M{(\choice(T_{i}))} & \text{if } A = \M T_{i}, \text{ for some } i = 1,\ldots,k.
\end{cases}
\end{equation}
Observe that, by the single-valuedness conditions present in $\varphi_{1}$, if $\M T_{i} = \M T_{j}$, for distinct $i$ and $j$, then $\M{(\choice(T_{i}))} = \M{(\choice(T_{j}))}$. Hence, the map $c$ is well-defined. In addition, the choice conditions yield that $c$ is indeed a total choice on $U$.

Let $\model = (U,M)$ be the set assignment differing from $\model_{1}$ only on the interpretation of the choice map symbol $\choice$, for which we have $\M \choice \defAs c$. Plainly, $\model$ coincides with $\model_{1}$ on all the choice-free terms in $\varphi$. In addition, since $c(\M T_{i}) = \M{(\choice(T_{i}))}$ for $i=1,\ldots,k$, the assignment $\model$ coincides with $\model_{1}$ on the remaining terms in $\varphi$ as well. Thus $\model \models \varphi$, proving that $\varphi$ is satisfiable when it admits a satisfiable $\BSTC^{-}$-reduction.

We have thus proved:
\begin{lemma}
Under unrestricted semantics, a complete $\BSTC$-formula in choice-flat form is satisfiable if and only if it admits a satisfiable $\BSTC^{-}$-reduction.\eod
\end{lemma}

In view of Theorem~\ref{NPcompl2LSS}, we can conclude: 
\begin{theorem}
Under unrestricted semantics, the satisfiability problem for $\BSTC$-formulae is \NP-complete.\eod
\end{theorem}

\subsubsection{$(\beta)$-semantics}
Let us now assume that our complete $\BSTC$-formula in choice-flat form $\varphi$ is satisfiable under the $(\beta)$-semantics, and let $\model=(U,M)$ be a model for it, where now $\M \choice$ is a choice satisfying the axiom $(\beta)$. Then the model $\model$ satisfies all the following instances of the axiom $(\beta)$:
\begin{description}
\item[$(\beta)$\emph{-conditions}:] $\big(T_{i} \subseteq T_{j} \: \wedge \: \choice(T_{i}) \cap \choice(T_{j}) \neq \emptyset \big) \;\; \Longrightarrow \;\; \choice(T_{i}) \subseteq \choice(T_{j})$, ~~for $i,j = 1,\ldots,k$.
\end{description}

Let $\varphi_{\beta}$ be the $\BSTC^{-}$-formula obtained by adding the $(\beta)$-conditions to the $\BSTC^{-}$-$\beta$-reduction of $\varphi$, while regarding the choice terms in it just as set variables (with no internal structure).  We call the formula $\varphi_{\beta}$ the $\BSTC^{-}$-\emph{$\beta$-reduction of $\varphi$}. Notice that $|\varphi_{\beta}| = \mathcal{O}(|\varphi|^{2})$. In addition, $\varphi_{\beta}$ is plainly satisfiable.

Conversely, let $\varphi_{\beta}$ be satisfiable and let $\model_{\beta} = (U,M_{\beta})$ be a model for $\varphi_{\beta}$. Let $\Omega_{\beta} \defAs \{\Mbeta{T_{i}} : i = 1, \ldots,k\}$ and $c_{\beta}$ be a map over $\Omega_{\beta}$ such that $c_{\beta}(\Mbeta{T_{i}}) =  \Mbeta{(\choice(T_{i}))}$, for $i = 1, \ldots,k$. From the choice and single-valuedness conditions, $c_{\beta}$ is a choice over the domain $\Omega_{\beta}$. In addition, by the $\beta$-conditions present in $\varphi_{\beta}$, the choice $c_{\beta}$ satisfies the $(\beta)$-axiom. Hence, Theorem~\ref{THM:lifting beta} yields that $c_{\beta}$ has the $(\beta)$-lifting property, i.e., there is a total choice $\cPlus_{\beta} \colon \powPlus \rightrightarrows U$ extending $c_{\beta}$ and satisfying the $(\beta)$-axiom. Let $\model = (U,M)$ be the set assignment differing from $\model_{\beta}$ only on the interpretation of the choice symbol $\choice$, for which we have $\M \choice = \cPlus_{\beta}$. It is not hard to check that $\model \models_{\beta} \varphi$.

Thus, we have:
\begin{lemma}
A complete $\BSTC$-formula in choice-flat form is satisfiable under the $(\beta)$-semantics if and only if it admits a satisfiable $\BSTC^{-}$-$\beta$-reduction.\eod
\end{lemma}

Since the size of the $\BSTC^{-}$-$\beta$-reduction of a given $\BSTC$-formula $\psi$ is at most quadratic in the size of $\psi$, in view of Theorem~\ref{NPcompl2LSS} we can conclude: 
\begin{theorem}
Under the $(\beta)$-semantics, the satisfiability problem for $\BSTC$-formulae is \NP-complete.\eod
\end{theorem}

\subsubsection{$(\alpha)$-semantics}
Finally, we assume that our complete $\BSTC$-formula in choice-flat form $\varphi$ is satisfiable under the $(\alpha)$-semantics. Let $\model=(U,M)$ be a model for it, where now $\M \choice$ is a choice satisfying the axiom $(\alpha)$. Then, the model $\model$ satisfies also the following instances of the axiom $(\alpha)$:
\begin{description}
\item[$(\alpha)$\emph{-conditions}:] $T_{i} \subseteq T_{j}  \;\; \Longrightarrow \;\; T_{i} \cap \choice(T_{j}) \subseteq \choice(T_{j})$, ~~for $i,j = 1,\ldots,k$.
\end{description}
In addition, let $\Omega \defAs \{\M{T_{i}} : i = 1, \ldots, k\}$. Then, plainly, the partial choice correspondence $\M\choice\restrict{\Omega}$ over the choice domain $\Omega$ has the $(\alpha)$-lifting property. Hence, by Theorem~\ref{THM:lifting alpha}\ref{c}, for every $\emptyset \neq \mathcal{B} \subseteq \Omega$ such that $\mathcal{B}$ is $\subseteq$-closed w.r.t.\ $\Omega$, we have
\begin{equation}\label{nonEmptynessSem}
\bigcup \mathcal{B} \setminus \bigcup\nolimits_{B \in \mathcal{B}} \overline{c}(B) \neq \emptyset\,.
\end{equation}
In fact, (\ref{nonEmptynessSem}) holds for every $\emptyset \neq \mathcal{B} \subseteq \Omega$, irrespectively of whether $\mathcal{B}$ is $\subseteq$-closed w.r.t.\ $\Omega$ or not, as can be easily checked. Thus, $\model$ satisfies also the following further conditions:
\begin{description}
\item[\emph{nonemptiness conditions}:] $\bigcup_{i \in I} T_{i} \setminus \bigcup_{i \in I}(T_{i} \setminus \choice(T_{i}))  \neq \varnothing$, ~~for each $\emptyset \neq I \subseteq \{1,\ldots,k\}$.
\end{description}

Let $\varphi_{\alpha}$ be the $\BSTC^{-}$-formula obtained by adding the $(\alpha)$- and the nonemptiness conditions to the $\BSTC^{-}$-reduction of $\varphi$, while regarding the choice terms in it just as set variables (with no internal structure).  We call the formula $\varphi_{\alpha}$ the $\BSTC^{-}$-\emph{$\alpha$-reduction of $\varphi$}. Notice that $|\varphi_{\alpha}| = \mathcal(|\varphi|^{2} + k\cdot 2^{k})$. In addition, $\varphi_{\alpha}$ is plainly satisfiable.

Conversely, let us assume that $\varphi_{\alpha}$ is satisfiable and let $\model_{\alpha} = (U,M_{\alpha})$ be a model for it. Let $\Omega_{\alpha} \defAs \{\Malpha{T_{i}}  : i = 1, \ldots,k\}$ and $c_{\alpha}$ be a map over $\Omega_{\alpha}$ such that $c_{\alpha}(\Malpha{T_{i}}) =  \Malpha{X_{i}}$, for $i = 1, \ldots,k$. As before, thanks to the choice and the single-valuedness conditions, $c_{\alpha}$ is a choice over the domain $\Omega_{\alpha}$. In addition, from the $(\alpha)$- and the nonemptiness conditions, Theorem~\ref{THM:lifting alpha} yields that $c_{\alpha}$ has the $(\alpha)$-lifting property, i.e., there is a total choice $\cPlus_{\alpha} \colon \powPlus \rightrightarrows U$ extending $c_{\alpha}$ and satisfying the $(\alpha)$-axiom. Let $\model = (U,M)$ be the set assignment differing from $\model_{\alpha}$ only on the interpretation of the choice symbol $\choice$, for which we have $\M \choice = \cPlus_{\alpha}$. It is routine to check that $\model \models_{\alpha} \varphi$.

Hence, we have:
\begin{lemma}
Under the $(\alpha)$-semantics, a complete $\BSTC$-formula in choice-flat form is satisfiable if and only if it admits a satisfiable $\BSTC^{-}$-$\alpha$-reduction.\eod
\end{lemma}

As observed earlier, the size of $\varphi_{\alpha}$ is $\mathcal{O}(|\varphi|^{2} + k\cdot 2^{k})$. Thus, in general, the satisfiability problem for $\BSTC$-formulae under the $(\alpha)$-semantics is in \NEXP. However, if the number of distinct choice terms is restricted to be $\mathcal{O}(1)$, we have $|\varphi_{\alpha}| = \mathcal{O}(|\varphi|^{2})$ and therefore, in view of Theorem~\ref{NPcompl2LSS}, we have:
\begin{theorem}
Under the $(\alpha)$-semantics, the satisfiability problem for $\BSTC$-formulae with $\mathcal{O}(1)$ distinct choice terms is \NP-complete.\eod
\end{theorem}

\subsubsection{Rationalizable-semantics}
We say that a $\BSTC$-formula $\varphi$ is satisfiable under Rationalizable-semantics if it admits a model $\mathcal{M} = (U,M)$ such that the interpreted
choice map $\choice^{M}$ is rationalizable by some relation $R$, in that case we write $M \models_{R} \varphi$.

Assume now that a $\BSTC$-formula in choice-flat form $\varphi$ is satisfiable under rationalizable-semantics. Let $\mathcal{C}_\varphi$ be the collection
 of all choice terms in $\varphi$ and $V_0$ be the collection of all individual variables in $\varphi$. Then put $\psi_\varphi$ has the conjunction:
\[
\psi_\varphi \coloneqq \bigwedge_{\choice(T) \in \mathcal{C}_\varphi}( T \setminus \choice(T) \subseteq \medcup T \wedge 
\choice(T) \cap \medcup T = \varnothing)\ \wedge\ \bigwedge_{x,y \in V_0} x \neq y.
\]
Of course in order to satisfy  $\psi_\varphi$ by a set assignment $(U,M)$, the elements of $U$ must be such that, the operator $\cup$, is defined between
them, as for the sets in the von Neumann universe of all sets $\mathcal{N}$.

Let $(U,M)$ be a set assignment for $\varphi$ under rationalizable-semantics. Assign to each individual variable $x$ a set $b_x$, so that for each 
pair $x,y$, $b_x = b_y \iff x = y$. By Theorem~\ref{THM:lifting rationalizable}, $\choice^{M}$ admits an acyclic selection $\pi$.
Put for each individual variable:
\[
(x)^{M'} \coloneqq \{y^{M'}\ |\ (x^{M},y^{M}) \in ran(\pi)\} \cup \{b_x\},
\]
and for each set variable:
\[
(X)^{M'} \coloneqq \{x^{M'}\ |\ x^{M} \in X^{M}\}.
\]
Notice that, since $ran(\pi)$ is acyclic, $(x)^{M'}$ is a well founded set for each individual, and also that 
since $b_x$ belongs only to $x^{M'}$, we have that:
\begin{equation}
x^{M} \in X^{M} \iff x^{M'} \in X^{M'}, \label{rationalizable model equiv}
\end{equation}
therefore $M' \models \bigwedge_{x,y \in V_0} x \neq y$.

We prove that $(\mathcal{N},M')$ is a model for$\varphi \wedge \psi_\varphi$.

First by $M \models \varphi$ and \eqref{rationalizable model equiv} we have $M' \models \varphi$.

Now let $y^{M'} \in T^{M'} \setminus (\choice(T))^{M'}$, then by \eqref{rationalizable model equiv}, $y^{M} \in T^{M} \setminus (\choice(T))^{M} = (\overline{\choice}(T))^{M}$.
Let $(x^{M},y^{M}) = \pi(y^{M},T^{M})$, then $x^{M} \in T^{M}$, thus $x^{M'} \in T^{M'}$, and $(x^{M},y^{M}) \in ran(\pi)$, therefore
$y^{M'} \in x^{M'}$, namely $M' \models \bigwedge_{\choice(T) \in \mathcal{C}_\varphi} T \setminus \choice(T) \subseteq \medcup T$.

Finally let $x^{M'} \in (\choice(T))^{M'}$, thus by \eqref{rationalizable model equiv} $x^{M} \in (\choice(T))^{M}$. We have that, for each $y^{M} \in T^{M}$, $(y^{M},x^{M})
\notin P$, where $P$ is defined as in Definition~\ref{DEF:rationalizable selection}, thus $x^{M'} \notin y^{M'}$, henceforth
$M' \models \bigwedge_{\choice(T) \in \mathcal{C}_\varphi} \choice(T) \cap \medcup T = \varnothing$.

Summing up $M'$ is a model for $\varphi \wedge \psi_\varphi$. So we can conclude that if $\varphi$ is satisfiable under rationalizable-semantics,
then $\varphi_R = \varphi \wedge \psi_\varphi$ is satisfiable under unrestricted semantics and so is is $\BSTC^{-}$-reduction $\varphi_{R}^{-}$.
\\

Conversely let $\varphi_{R}^{-}$, the $\BSTC^{-}$-reduction of $\varphi_R = \varphi \wedge \psi_\varphi$, where $\varphi$ is any $\BSTC$-formula, be satisfiable by some
set assignment $(\mathcal{N},M)$. Let $U$ be a collection of elements and let $f : \mathcal{N} \longrightarrow U$ be injective,
then define the set assignment $(U,M')$ as:
\[
x^{M'} \coloneqq f(x^{M}),\quad X^{M'} \coloneqq \{f(x^{M})\ |\ x^{M} \in X^{M'}\},
\]
where $x$ is an individual variable and $X$ is a set variable.

Since $M \models x \neq y$, for all $x,y \in V_0$, we have that \eqref{rationalizable model equiv} still holds.

Since $M \models \varphi_{R}^{-}$ it is easy to prove that $M' \models \varphi^{-}$, that is the $\BSTC^{-}$-reduction of $\varphi$. Thanks to the choice and 
single-valuedness conditions, the map $\choice_{\varphi}(T^{M'}) = (\choice(T))^{M'}$ is a choice correspondence over the domain
$\Omega \coloneqq \{T\ |\ T $ is a choice term of $\varphi\}$.

Let $P_{\varphi} = \{(x^{M'},y^{M'})\ |\ y^{M} \in x^{M}\}$. For any pair $(x^{M'},y^{M'}) \in P_{\varphi}$,
we have that $x^{M'} \neq y^{M'}$, otherwise $x^{M} \in x^{M}$. If there would exists a $T^{M'} \in \Omega$
such that $x^{M'},y^{M'} \in T^{M'}$, then $y^{M'} \notin (\choice(T))^{M'} = \choice_{\varphi}(T^{M'})$,
otherwise we would have $y^{M} \in (\choice(T))^{M}$, and $y^{M} \in x^{M} \subseteq \medcup T^{M}$ a contradiction 
since $M \models \psi_\varphi$. Thus $(x^{M'},y^{M'}) \in P$, where $P$ is the same as in
Definition~\ref{DEF:rationalizable selection} with regards to $\choice_{\varphi}$.

Furthermore $P_{\varphi}$ is acyclic, in fact if there would be a cycle $(x_0^{M'},x_1^{M'}),(x_1^{M'},x_2^{M'}),\ldots,(x_n^{M'},x_0^{M'})$
then $x_0^{M} \in x_1^{M} \in \ldots \in x_n^{M} \in x_0^{M}$.

Put $P_{\varphi}(y^{M'},T^{M'}) = P(y^{M'},T^{M'}) \cap P_{\varphi}$,
and $\oc_\varphi(T^{M'}) = T^{M'} \setminus \choice_\varphi(T^{M'})$, then for each $y^{M'} \in \oc_{\varphi}(T^{M'})$, 
$P_{\varphi}(y^{M'},T^{M'})$ is non empty. In fact $M \models T \setminus \choice(T) \subseteq \medcup T$, meaning that for every $y^{M'} \in T^{M'}$,
there exists a $x^{M'} \in T^{M'}$ such that $y^{M} \in x^{M}$.
Thus any function $\pi$ with domain $\{(y^{M'},T^{M'})| T^{M'} \in \Omega, y^{M'} \in \oc_\varphi(T^{M'})\}$ such that $\pi(y^{M'},T^{M'})
\in P_{\varphi}(y^{M'},T^{M'})$ is an acyclic selection of $\choice_\varphi$.

Summing up, by Theorem~\ref{THM:lifting rationalizable} there exists a choice $\cPlus$ such that $\choice_\varphi =\cPlus\restrict{\Omega}$, therefore
$\varphi$ is satisfiable under Rationalizable-semantics.

In conclusion notice that the size of $\psi_\varphi$ is double the number $k$ of choice terms in $\varphi$ thus the size of $\varphi_{R}^{-}$ is 
$\Theta(|\varphi| +2k)$, and since $\varphi_{R}^{-}$ is a $MLS$ formula we can conclude that:
\begin{theorem}
Under Rationalizable-semantics, the satisfiability problem for $\BSTC$-formulae is \NP-complete.
\end{theorem}

\section{Conclusions} \label{SECT:Conclusions}

In this paper we have initiated the study of the satisfiability problem for unquantified formulae of an elementary fragment of set theory enriched with a choice correspondence symbol. 
Apart from the obvious theoretical reasons that motivate this approach, our analysis has its roots in applications within the field of choice theory.
In fact, the satisfiability tests implicit in our results naturally yield an effective way of checking whether the observed choice behavior of an economic agent is induced by an underlying rationality on the set of alternatives.

Future research on the topic is related to the extension of the current approach to a more general setting. 
In this direction, it is natural to examine the satisfiability problem for semantics characterized by other types of axioms of choice consistency, which are connected to rationalizability issues.
We are currently studying the lifting (and the associated satisfiability problem) of several combinations of axioms, namely, $(\alpha)$ adjoined with $(\gamma)$ and/or $(\rho)$ (see~\cite{CanGiaWat17}). 
The motivation of this analysis is that the satisfaction of these axioms is connected to the (quasi-transitive) rationalizability of a choice. 
More generally, it appears natural to examine the lifting of $(m,n)$\textit{-Ferrers properties} in the sense of \cite{GiaWat14} (see also~\cite{GiaWat17}).
In fact, these properties give rise to additional types of rationalizability -- the so-called $(m,n)$\textit{-rationalizability} -- in which the relation of revealed preference satisfies structural forms of pseudo-transitivity (see~\cite{CanGiaGreWat16}).

\newcommand{\TLQSTR}{\ensuremath{\mbox{$\mathrm{3LQST_{0}^{R}}$}}\xspace}
\newcommand{\QLQSR}{\ensuremath{\mbox{$4\mathrm{LQS}^{R}$}}\xspace}
We also intend to find decidable extensions with choice correspondence terms of the three-sorted fragment of set theory $\TLQSTR$ (see \cite{CN16}) and of the four-sorted fragments $(\QLQSR)^h$, with $h \in \mathbb{N}$ (see \cite{CanNic13a}), which admit a restricted form of quantification over individual and set variables (in the case of $\TLQSTR$), and also over collection variables (in the case of $(\QLQSR)^h$. The resulting decision procedures would allow to reason automatically on very expressive properties in choice theory.

\bibliographystyle{eptcs}

\end{document}